\documentclass[twocolumn,trackchanges,floatfix]{aastex631}

\usepackage{amsmath}
\usepackage{amssymb}

\received{}
\revised{}
\accepted{}
\submitjournal{}

\shorttitle{NIRSpec TSO Commissioning Performance}

\shortauthors{Espinoza et al.}

\graphicspath{{./}{figures/}}

\begin{document}

\title{Spectroscopic Time-series Performance of JWST/NIRSpec from Commissioning Observations}

\correspondingauthor{N\'estor Espinoza}
\email{nespinoza@stsci.edu}

\author[0000-0001-9513-1449]{N\'estor Espinoza}
\affiliation{Space Telescope Science Institute, 3700 San Martin Drive, Baltimore, MD 21218, USA}
\affiliation{Department of Physics \& Astronomy, Johns Hopkins University, Baltimore, MD 21218, USA}

\author[0000-0001-7130-2880]{Leonardo \'Ubeda}
\affiliation{Space Telescope Science Institute, 3700 San Martin Drive, Baltimore, MD 21218, USA}

\author[0000-0001-7058-1726]{Stephan M. Birkmann}
\affiliation{European Space Agency, c/o STScI, 3700 San Martin Drive, Baltimore, USA}

\author[0000-0001-8895-0606]{Pierre Ferruit}
\affiliation{European Space Agency, ESAC, 28692 Villafranca del Castillo, Madrid, Spain}

\author[0000-0003-3305-6281]{Jeff A. Valenti}
\affiliation{Space Telescope Science Institute, 3700 San Martin Drive, Baltimore, MD 21218, USA}

\author[0000-0001-6050-7645]{David K. Sing}
\affiliation{Department of Physics and Astronomy, Johns Hopkins University, Baltimore, MD 21218, USA}
\affiliation{Department of Earth \& Planetary Sciences, Johns Hopkins University, Baltimore, MD, USA}

\author{Zafar Rustamkulov}
\affiliation{Department of Earth \& Planetary Sciences, Johns Hopkins University, Baltimore, MD, USA}

\author{Michael Regan}
\affiliation{Space Telescope Science Institute, 3700 San Martin Drive, Baltimore, MD 21218, USA}

\author{Sarah Kendrew}
\affiliation{European Space Agency, c/o STScI, 3700 San Martin Drive, Baltimore, USA}

\author[0000-0003-2954-7643]{Elena Sabbi}
\affiliation{Space Telescope Science Institute, 3700 San Martin Drive, Baltimore, MD 21218, USA}

\author[0000-0001-8291-6490]{Everett Schlawin}
\affiliation{Steward Observatory, 933 North Cherry Avenue, Tucson, AZ 85721, USA}

\author{Thomas Beatty}
\affiliation{Steward Observatory, 933 North Cherry Avenue, Tucson, AZ 85721, USA}

\author{Lo\"ic Albert}
\affiliation{Institut de recherche sur les exoplan\`etes and D\'epartement de Physique, Universit\`e de Montr\'eal, 1375 Avenue Th\'er\`ese-Lavoie-Roux, Montr\'eal, QC, H2V 0B3, Canada}

\author{Thomas P. Greene}
\affiliation{NASA Ames Research Center, Space Science and Astrobiology Division, M.S. 245-6, Moffett Field, CA 94035, USA}

\author{Nikolay Nikolov}
\affiliation{Space Telescope Science Institute, 3700 San Martin Drive, Baltimore, MD 21218, USA}

\author{Diane Karakla}
\affiliation{Space Telescope Science Institute, 3700 San Martin Drive, Baltimore, MD 21218, USA}

\author{Charles Keyes}
\affiliation{Space Telescope Science Institute, 3700 San Martin Drive, Baltimore, MD 21218, USA}

\author[0000-0002-5320-2568]{Nimisha Kumari}
\affiliation{AURA for the European Space Agency, ESA Office, Space Telescope Science Institute, 3700 San Martin Drive, Baltimore, MD, 21218 USA}

\author[0000-0003-2896-4138]{Catarina Alves de Oliveira}
\affiliation{European Space Agency, ESAC, 28692 Villafranca del Castillo, Madrid, Spain}

\author[0000-0002-5666-7782]{Torsten Böker}
\affiliation{European Space Agency, c/o STScI, 3700 San Martin Drive, Baltimore, USA}

\author{Maria Pena-Guerrero}
\affiliation{Space Telescope Science Institute, 3700 San Martin Drive, Baltimore, MD 21218, USA}

\author[0000-0002-9262-7155]{Giovanna Giardino}
\affiliation{ATG Europe for the European Space Agency, ESTEC, Keplerlaan 1, Noordwijk, The Netherlands}

\author[0000-0003-0192-6887]{Elena Manjavacas}
\affiliation{AURA for the European Space Agency, ESA Office, Space Telescope Science Institute, 3700 San Martin Drive, Baltimore, MD, 21218 USA}

\author[0000-0001-7617-5665]{Charles Proffitt}
\affiliation{Space Telescope Science Institute, 3700 San Martin Drive, Baltimore, MD 21218, USA}

\author[0000-0002-7028-5588]{Timothy Rawle}
\affiliation{European Space Agency, c/o STScI, 3700 San Martin Drive, Baltimore, USA}

\begin{abstract}

We report on \emph{JWST} commissioning observations of the transiting 
exoplanet HAT-P-14~b, obtained using the Bright Object Time Series (BOTS) mode of the NIRSpec instrument with the G395H/F290LP grating/filter combination ($3-5\mu$m). While the data were used primarily to verify that the NIRSpec BOTS mode is working as expected, and to enable it 
for general scientific use, they yield a precise transmission spectrum which we find 
is featureless down to the precision level of the instrument, consistent with expectations given HAT-P-14~b's small scale-height and hence expected atmospheric features. The exquisite quality and stability 
of the \emph{JWST/NIRSpec} transit spectrum --- almost devoid of any systematic 
effects  --- allowed us to obtain median uncertainties of 50-60 ppm in this wavelength range at a resolution of $R=100$ in a single exposure, 
which is in excellent agreement with pre-flight expectations and close to the (or at the) photon-noise limit for a $J = 9.094$, F-type star like HAT-P-14. These observations showcase the ability of 
NIRSpec/BOTS to perform cutting-edge transiting exoplanet atmospheric science, 
setting the stage for observations and discoveries to be made in Cycle 1 and 
beyond.
\end{abstract}


\section{Introduction}

The NIRSpec/BOTS mode is one of the prime modes for transiting 
exoplanet science onboard the \emph{James Webb Space Telescope (JWST)} \citep{beichman, birkmann}. 
The mode offers precise spectroscopy of transiting exoplanets from 0.6 to 5 $\mu$m in a single exposure 
via its low resolution ($R\sim 100$) Prism mode, as well as high resolution (up to $R\sim 2700$) measurements using various combinations of dispersers and filters. As such, the NIRSpec instrument has unique capabilities 
to cover a wide range of science cases, and is set to perform observations 
of exoplanets of all sizes and temperature regimes, including worlds that could 
host suitable conditions for life as we know it \citep[see, e.g.,][]{lewis:trappist, neat:2017, rathcke:2021, lim:2021}.

During the commissioning of \emph{JWST}, Time Series Observations (TSOs) were 
obtained in all instruments to determine two key properties of the science instruments and modes. 
The first was to verify that these technically challenging observations were executing as planned. The 
second objective was to determine if the various instrument modes could be calibrated with sufficient accuracy
to precisely measure the small flux variations caused by the transits, and to identify additional calibration needs if limitations were found. 

For the \emph{JWST} instruments offering observations in the 
near-infrared (NIR; here defined as wavelengths up to 5 $\mu$m), a common 
target was decided to study the above mentioned properties of TSOs: the 
transiting exoplanet HAT-P-14~b \citep{torres:2010}. HAT-P-14~b is a dense 
($2.3M_J$; $1.15R_J$), short-period (4.6-day) exoplanet orbiting a relatively 
bright $J=9.09$, $1.5R_\odot$, low-activity F-star \citep{bonomo}. Given its massive nature, it 
has a relatively small scale-height ($H\sim 150$ km) which, combined with the 
large stellar radius, should give rise to small variations in the transit depth 
as a function of wavelength due to the exoplanetary atmosphere (20-60 ppm 
depending on the assumptions used to calculate this signature). This provided us 
with an excellent target to commission the NIRSpec/BOTS mode: a target for which we expect, based on reasonable assumptions, a featureless transmission spectrum down to the precision level of the instrument for a single transit. Any observed variations in the transit depth as a function of wavelength therefore would most likely be due to instrumental rather than astrophysical effects, and thus would allow us to pinpoint any irregularities in either the instrument and/or the data reduction and calibration. The HAT-P-14~b system also has the advantage of being thoroughly characterized via precise ground and space-based photometry \citep[see, e.g.,][]{saha, fukui, Simpson}, radial-velocities \citep[e.g.,][]{bonomo,torres:2010} and even adaptive optics and 
astrometric constraints on possible nearby companions \citep{belokurov, ngo}, which enabled us to identify 
causes for possible deviations from any solutions we obtained 
by analyzing \emph{JWST/NIRSpec} spectrophotometry.

Here we present the analysis and results of \emph{JWST} commissioning 
observations of a single transit event of the exoplanet HAT-P-14~b 
using the NIRSpec/BOTS mode; the very same data that were used to enable 
this mode for scientific use. The article is organized as follows. In 
Section \ref{sec:obs-dr}, we present the observations and data reduction of 
the dataset. In Section \ref{sec:tso} we present a detailed analysis of the TSO, 
along with performance metrics for the mode retrieved from these observations. 
Finally, we conclude with a discussion of our findings in Section \ref{sec:discussion} and a summary of our main findings in Section \ref{sec:summary}.

\section{Observations and Data Reduction}
\label{sec:obs-dr}
\subsection{Observations}
\label{sec:observations}

A TSO was obtained during the \emph{JWST} commissioning campaign on 
2022-05-30 between 00:30 and 07:02 UT with the NIRSpec/BOTS mode, 
targeting the star HAT-P-14 (PID 1118; PI: Proffitt). The objective of this observation was 
to measure the transit event of the exoplanet HAT-P-14~b in order 
to measure its transmission spectrum, i.e., the transit depth as a 
function of wavelength. The 6-hour exposure consisted of 1139 
integrations with 20 NRSRAPID groups per integration, taken with 
the G395H/F290LP grating/filter, which covers the wavelength range from 2.87 to 5.14 $\mu$m with a resolution of about $R\sim 2,700$, using the S1600A1 aperture and 
the SUB2048 subarray. This resulted in data gathered by both the NRS1 and NRS2 detectors, each having a height of 32 pixels and a width of 2048 pixels, which included four columns of reference pixels at the left and right edges of the selected subarray. The pixel scale, for reference, is about 100 milli-arcseconds (mas) per pixel. 

\begin{figure*}
\centering
\includegraphics[width=7.3in]{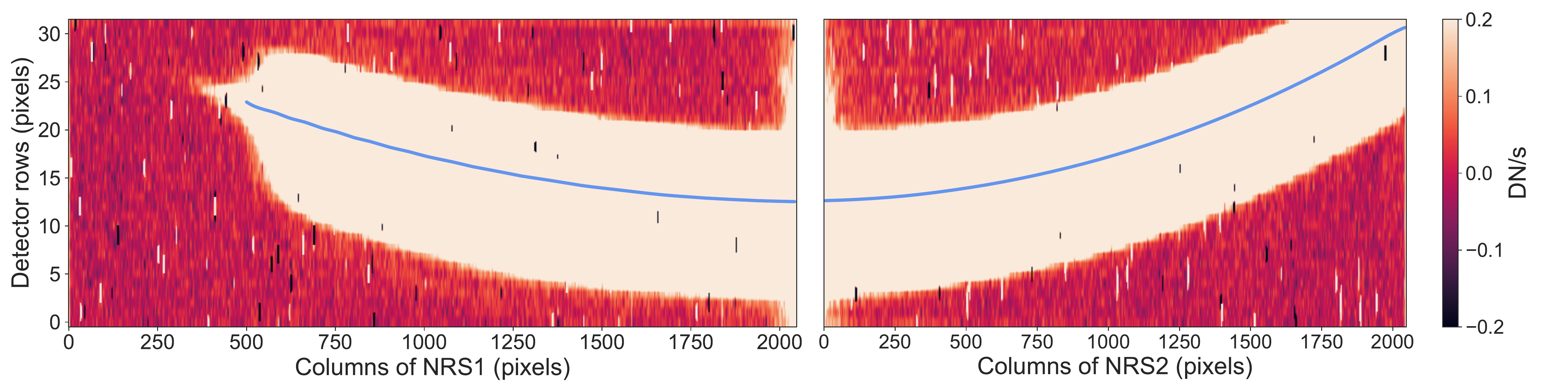}
\caption{Median signal of all integrations showcasing the 2D spectrum of HAT-P-14 as obtained by NIRSpec/BOTS G395H; the trace of the very first integration as obtained by our methods is shown in blue. These were obtained using the raw rates per integration produced by the JWST pipeline, for which the median background has been substracted. Hot and bad pixels in the NRS1 and NRS2 detectors are clearly observed, as well as what appears to be some scattered light at the right-edge of NRS1 and left edge of NRS2 (see text for details). Note that the countrate illustrated in the frame is significantly constrained (between -0.2 and 0.2 DN/s) in order to highlight the (minimum) background structure. Also note the aspect ratio of the frame has been stretched in particular in the 32-pixel cross-dispersion direction for illustration purposes.}\label{fig:2dspectrum}
\end{figure*}

Figure \ref{fig:2dspectrum} shows the median rates per integration 
across the entire exposure, which shows several interesting features. The first and most evident is the clear distorted shape of 
the spectral trace (blue curve; see Section \ref{sec:tracing} for details on how this was obtained). This covers from 2.7 to 
3.7 microns in NRS1, and from about 3.8 to 5 microns on NRS2. The second are a number of bad and hot pixels in the frame. Bad pixel masks are being provided by the NIRSpec team to identify those via the Calibration Reference Data System (CRDS)\footnote{\url{https://jwst-crds.stsci.edu/}}. The third feature is the lack of significant structure in the background, which suggests it will be relatively straightforward to remove it from observations. Finally, we also note what appears to be some scattered light evident on the right-most $\sim 50$ pixels in the NRS1 detector, which is also seen on a few of the left-most pixels of the NRS2 detector. The scale of the count rate in Figure \ref{fig:2dspectrum} makes this feature appear much more dramatic than it actually is in reality (by design --- we wanted to highlight this in the figure): the extra countrate of these enlarged wings only adds of the order of $\sim 1$ DN/s/pixel, whereas the peak counts in that region are about $\sim 1,000$ DN/s/pixel for HAT-P-14. Even if this component contained scattered light from nearby wavelengths, the resulting dilution of the extracted spectrum would be negligible.

It is important to note that the distortion of the right edge of the spectrum seen in the NRS2 detector made the spectra somewhat hard to extract, as it falls right on the corner of 
the detector. Based on these commissioning observations, the NIRSpec team therefore
decided to move the NRS2 subarray by four pixels in the vertical direction such that the 
spectrum is fully contained in the subarray, and can be more easily extracted. Therefore, the science observations obtained
after commissioning will have a slight offset between the 
spectral trace seen in NRS1 and NRS2.

\subsection{Spectral tracing}
\label{sec:tracing}

We describe the data reduction in detail below, starting from the rates per integration, i.e., 
the \texttt{rateint.fits} products, after reducing the \texttt{uncal.fits} files available via the Mikulski Archive for Space Telescopes (MAST\footnote{\url{https://archive.stsci.edu/}}) using version 1.6.2 of the JWST pipeline\footnote{\url{https://github.com/spacetelescope/jwst}} with its default parameters (which we note does not include sky substraction/calibration for this mode/instrument --- we detail how we deal with this in Section \ref{sec:1f}). 
From these products, we use the \texttt{transitspectroscopy} library which contains custom routines designed for transit spectroscopy  
measurements to reduce the data, which is publicly available via GitHub\footnote{\url{https://github.com/nespinoza/transitspectroscopy}}. First, we independently trace the spectral shape for each integration 
and for each detector by cross-correlating a Gaussian with each column, 
and finding the maximum of the resulting cross-correlation function 
(CCF). For the NRS1 detector, we start this process from column 500 
and for the NRS2 detector, we start the process from column 5, in 
both cases following the trace all the way through the right-edge 
of the detector. Outliers due to cosmic-rays, bad and hot 
pixels were identified on these trace shape measurements by running 
a median filter through the trace at each integration with a 
window of 11 pixels in the wavelength direction and finding trace positions 
deviating more than 5 standard deviations from this trace. The outlier-corrected traces 
were then smoothed using a spline. 

\begin{figure}
\centering
\includegraphics[width=3.3in]{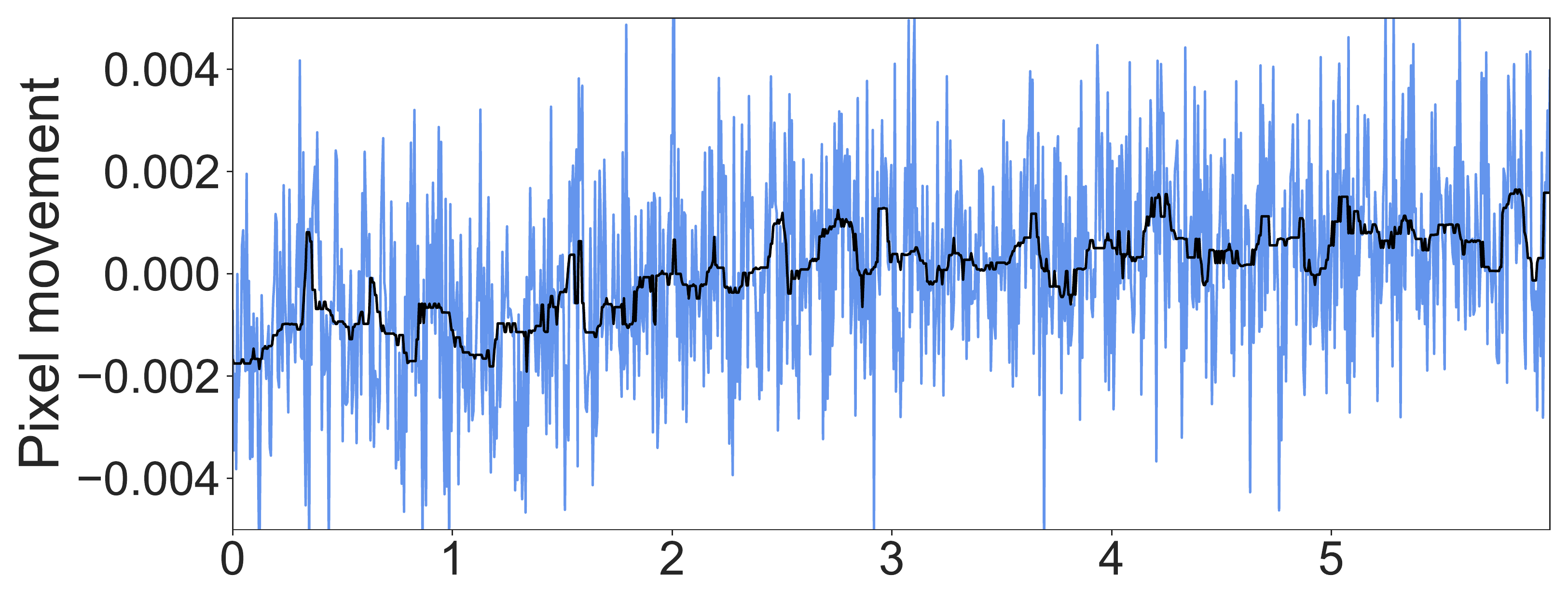}
\includegraphics[width=3.3in]{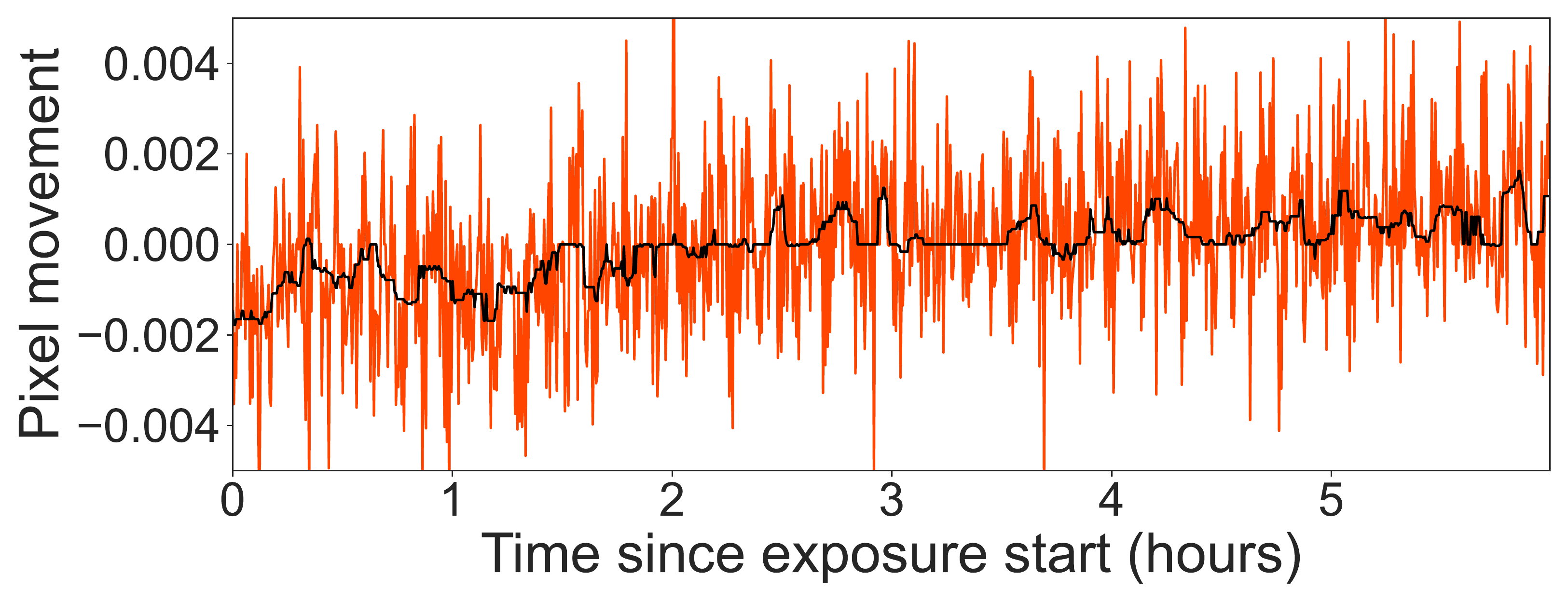}
\caption{Trace position movement for NRS1 (top, blue) and 
NRS2 (bottom, red) portions of the spectral trace as a function 
of time, as tracked by the median movement of all the points in the 
trace. Black line is a median filter with a window of 11 integrations 
through the measurements. Note the remarkable stability of the 
trace position, which stays on the same pixel down to 1/500th 
of a pixel (0.2 mas).}\label{fig:traces}
\end{figure}

The trace positions as a function of time are presented in Figure 
\ref{fig:traces}. These show remarkable stability throughout the 
exposure, even during a high-gain antenna (HGA) move that happened at about half an hour after the start of the exposure, showing deviations within 1/500th of a pixel (0.2 mas) over the 
6-hour exposure. There seems to be both high-frequency variations and a 
slight systematic movement of the trace during the first two hours, but these do not 
appear to cause any evident systematic trend on the actual spectrophotometry, and seem 
to be consistent across the two detectors. A detailed analysis of the trace time variability is given in Appendix \ref{appendix:trace}.

\subsection{1/f corrections}
\label{sec:1f}

1/f noise is an important component of the near-infrared 
\emph{JWST} detector noise which can give rise to significant 
scatter in a TSO if not accounted for and at least partially corrected \citep[see, e.g.,][]{schlawin, zafar}. We perform 1/f (and sky background) corrections at 
the integration level by first masking all pixels within 15 pixels 
from the traces, and taking the median of the non-masked pixels at 
each column, subtracting that from each pixel in that column. This methodology is, 
of course, not perfect because it doesn't take care of the intra-column component 
of 1/f noise; however, it has the added benefit of correcting for any sky background 
signal, as well as any column-to-column offset created by 1/f noise. We 
find that this correction significantly improves the 
precision of our analysis, as already suggested by the works of 
\cite{schlawin} and \cite{zafar}.

\subsection{Spectral extraction}
\label{sec:extraction}

Using the 1/f-corrected integrations and the traces obtained as described in the previous section, we proceed to extract the spectra at 
each integration. Note this implies we are not flat-fielding our data prior to extraction; the impact of which is pressumed to be small given the precise stability of the spectrum presented in the previous sections. Given that we find a number of residual cosmic rays, as well as bad and hot pixels not 
corrected by the JWST STScI pipeline, 
we decided to extract the spectrum using the optimal extraction 
algorithm for distorted spectra of \cite{Marsh}, which automatically 
takes care of outliers in the 2D spectral profile. The implementation 
we use is the one described in \cite{Brahm:2017}, with the difference 
that we use a variance array as input to 
the algorithm. 

\begin{figure*}
\centering
\includegraphics[width=7in]{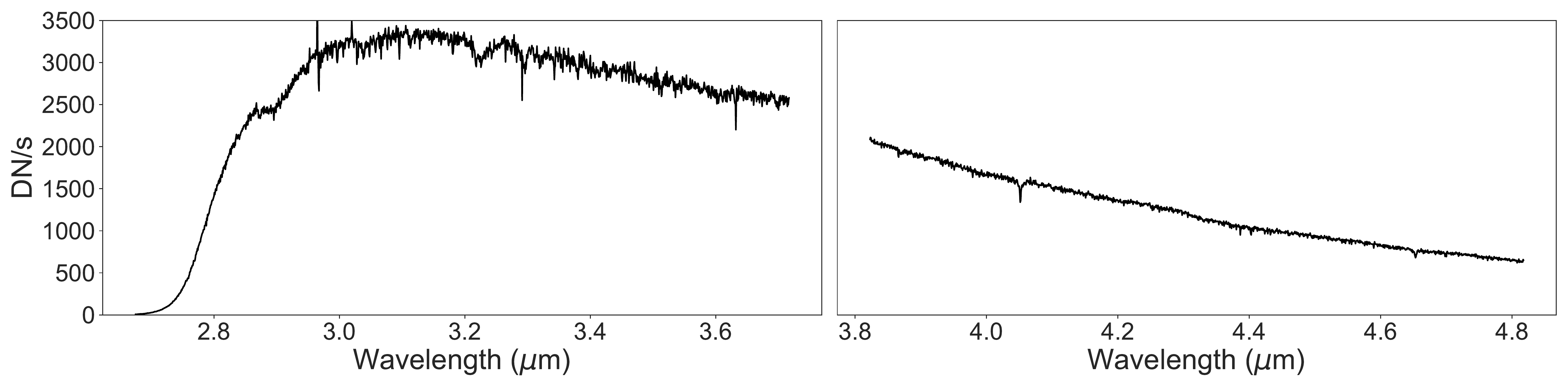}
\caption{Median extracted spectrum for HAT-P-14 using optimal extraction on the frame shown in Figure \ref{fig:2dspectrum} for NRS1 (left) and NRS2 (right). Note we only extract up to 4.8 $\mu$m as the rest of the spectral range up to 5 $\mu$m falls at the very corner of the detector; this will not be the case for future science observations as the subarray has since been moved by 4 pixels to avoid this (see text for details). Also note the spectra has not been flat fielded, which explains the high frequency variations as a function of wavelength especially in the NRS1 spectrum.}\label{fig:1dspectrum}
\end{figure*}

The spectrum is extracted using a 14-pixel aperture around the trace 
for NRS1 and NRS2 independently; this aperture size was selected as to be consistent across the 
extracted wavelength range, as well as to be able to maximize the signal-to-noise ratio while 
being able to extract the spectrum from as a wide wavelength range as possible (other smaller and 
larger apertures gave overall similar results). For NRS2, 
we don't extract the spectrum all the way to the corner; instead, our chosen aperture only allows us 
to extract the spectrum up to 4.8 microns. Wavelengths 
are assigned to each column by making use of the JWST pipeline's 
wavelength map, which is in turn obtained through the NIRSpec instrument model, which was observed 
during commissioning to work according to pre-flight specifications, which are excellent for the purposes 
of this work \citep{nirspec:nl}. The extracted median 
spectrum of HAT-P-14 is shown in Figure \ref{fig:1dspectrum}. As can 
be observed, only a small fraction of the large number of bad and 
hot pixels clearly seen in Figure \ref{fig:2dspectrum} remain after using our 
optimal extraction procedure.

Using these extracted spectra, we then move to a discussion on the 
analysis and TSO observed performance during commissioning in the next 
section.

\section{TSO analysis and performance}
\label{sec:tso}

\subsection{Band-integrated light curve analysis}
\label{sec:white-light-analysis}

\begin{figure*}
\centering
\includegraphics[width=7in]{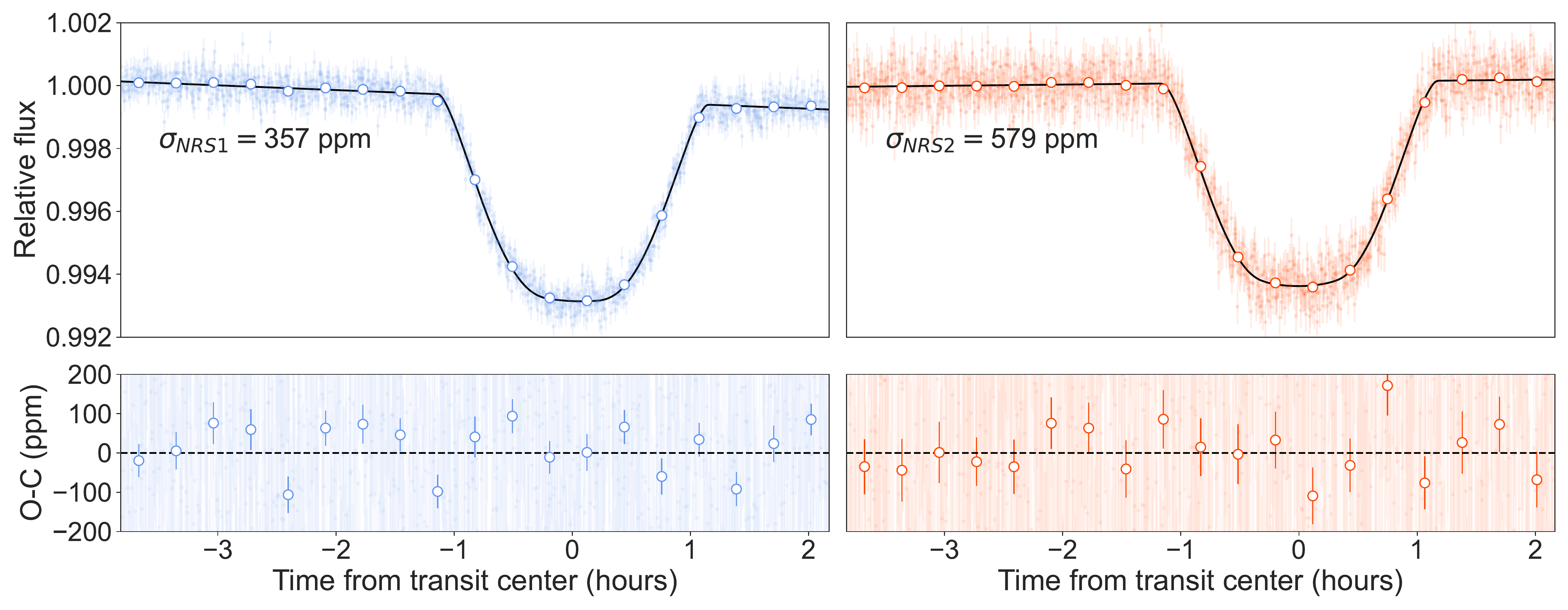}
\caption{band-integrated transit light curves for HAT-P-14~b as observed by NIRSpec/BOTS G395H via the NRS1 (left) and NRS2 (right) detectors. These have been fitted simply by a transit model and a linear trend, which seems to be sufficient to match the variations observed in the data. Note the quoted precision is at a cadence of 19 seconds after fitting.}\label{fig:transit}
\end{figure*}

As a first-order analysis, we construct the ``band-integrated" light curve of 
HAT-P-14~b by adding up the light from NRS1 and NRS2 separately. We decided to fit 
these light curves with a simple model which included a \texttt{batman} 
transit model \citep{batman} and a linear trend in time\footnote{More complex models were tried as well (e.g., a few models including this trend plus a Gaussian Process); however, all of those models were indistinguishable with this simple one judging from the bayesian evidence of our fits ($|\Delta \ln \mathcal{Z}| \lesssim 2$).}. We used 
a square-root law to parametrize limb-darkening in these light curve fits as, 
based on simulations performed by our team similar to those suggested 
in \cite{ldlaws}, this was one of the laws that performed the best to 
precisely and accurately recover input transit parameters. As priors 
for our fit, we used orbital parameters retrieved by fitting \emph{TESS} 
transit light curves from Sectors 25 and 26 in the exact same manner as 
described in \cite{patel}, from which we obtained $a/R_* = 8.34 \pm 0.31$ 
and $b = 0.90 \pm 0.02$, and for which we left the eccentricity and 
argument of periastron fixed to the values found in the literature 
\cite[][$e=0.11$; $\omega=106.1$ deg]{bonomo} --- which we also fixed 
in our JWST/NIRSpec light curve fits. Finally, we used the parametrization of 
\cite{kipping} in our band-integrated light curve fits, leaving the transformed 
limb-darkening coefficients $(q_1, q_2)$ as free parameters in our fits with 
uniform priors between 0 and 1. A jitter term was also added and fitted 
to both light curves. The \texttt{juliet} \citep{juliet} library was used to 
perform these light curve fits independently for NRS1 and NRS2, using 
\texttt{dynesty} as the sampler via dynamic nested sampling 
\citep{dynesty}. The results of those fits are presented in Figure 
\ref{fig:transit}; a Table with our priors and posteriors is presented 
in Table \ref{tab:wl-params}.

\begin{deluxetable*}{lcccc}[t]
\tablecaption{Prior and posterior parameters of the band-integrated lightcurve fits performed on the NRS1 and NRS2 data of HAT-P-14~b. For the priors, $N(\mu,\sigma^2)$ stands for a normal distribution with mean $\mu$ 
and variance $\sigma^2$; $U(a,b)$ stands for a uniform distribution between $a$ and $b$, respectively and $\log U(a,b)$ stands for a log-uniform prior on the same ranges. Priors for $b$ and $a/R_*$ come from \textit{TESS} (see text for details).\label{tab:wl-params}}
\tablecolumns{3}
\tablewidth{0pt}
\tablehead{
\colhead{Parameter} &
\colhead{Prior} &
\colhead{Posterior (NRS1)} & 
\colhead{Posterior (NRS2)} & 
\colhead{Posterior (combined)}
}
\startdata
\multicolumn{3}{l}{\textit{Physical \& orbital parameters}} \\
\ \ \ \ $P$ (days) & $4.62767$ (fixed)  &  &  & \\
\ \ \ \ $t_{0}$ (BJD TDB) &  $N(2459729.71,0.2^2)$ &  $2459729.70662^{+0.00010}_{-0.00010}$ & $2459729.70696^{+0.00016}_{-0.00017}$ & $2459729.706791^{+0.000094}_{-0.000093}$ \\
\ \ \ \ $R_{p}/R_\star^{\psi}$  &  $U(0.0,0.2)$ &  $0.0826^{+0.0020}_{-0.0014}$  & $0.0849^{+0.0022}_{-0.0018}$ & $0.0839^{+0.0014}_{-0.0012}$\\
\ \ \ \ $b=(a/R_\star)\cos(i)$  &  $N(0.90,0.02^2)$ &  $0.9081^{+0.0049}_{-0.010}$ & $0.8966^{+0.0068}_{-0.0084}$ & $0.9016^{+0.0048}_{-0.0060}$\\
\ \ \ \ $a/R_\star$  &  $N(8.34,0.31^2)$ &  $8.29^{+0.18}_{-0.11}$ & $8.40^{+0.17}_{-0.15}$ & $8.354^{+0.12}_{-0.098}$\\
\ \ \ \ $e$  &  fixed &  0.11 & 0.11 & \\
\ \ \ \ $\omega$  &  fixed & 106.1 & 106.1 & \\
\multicolumn{3}{l}{\textit{Limb-darkening coefficients}} \\
\ \ \ \ $q_{1}^\dagger$ &  $U(0,1)$ &  $0.14^{+0.41}_{-0.11}$ & $0.49^{+0.29}_{-0.28}$ & \\
\ \ \ \ $q_{2}^\dagger$ &  $U(0,1)$ &  $0.58^{+0.30}_{-0.40}$ & $0.58^{+0.28}_{-0.34}$ & \\
\multicolumn{3}{l}{\textit{Instrumental parameters}} \\
\ \ \ \ $M_\textrm{flux}$ (ppm)&  $U(0,10^5)$ & $313^{+13}_{-13}$ & $-77^{+21}_{-22}$ & \\
\ \ \ \ $\theta_0^\ddag$ (ppm)&  $U(-10,10)$ & $-258^{+11}_{-10}$ & $66^{+18}_{-18}$ & \\
\ \ \ \ $\sigma_w$ (ppm) &  $\log U(10,10^4)$ &  $336.8^{+7.4}_{-7.8}$ & $534.8^{+12.4}_{-11.8}$ & \\
\enddata
\tablenotetext{\psi}{For a more precise estimate on the transit depth, see Figure \ref{fig:tspec}. See text for an explanation of why the band-integrated light curve achieves lower precisions than the light curves at each resolution element sampled by the instrument.}
\tablenotetext{\dagger}{These parameterize the square-root limb-darkening law using the transformations in 
\cite{kipping}.}
\tablenotetext{\ddag}{The linear trend slope was fitted to the standardized times, i.e., to the mean-subtracted time-stamps divided by the standard deviation of the time-stamps.}
\end{deluxetable*}

\begin{figure}
\centering
\includegraphics[width=3.3in]{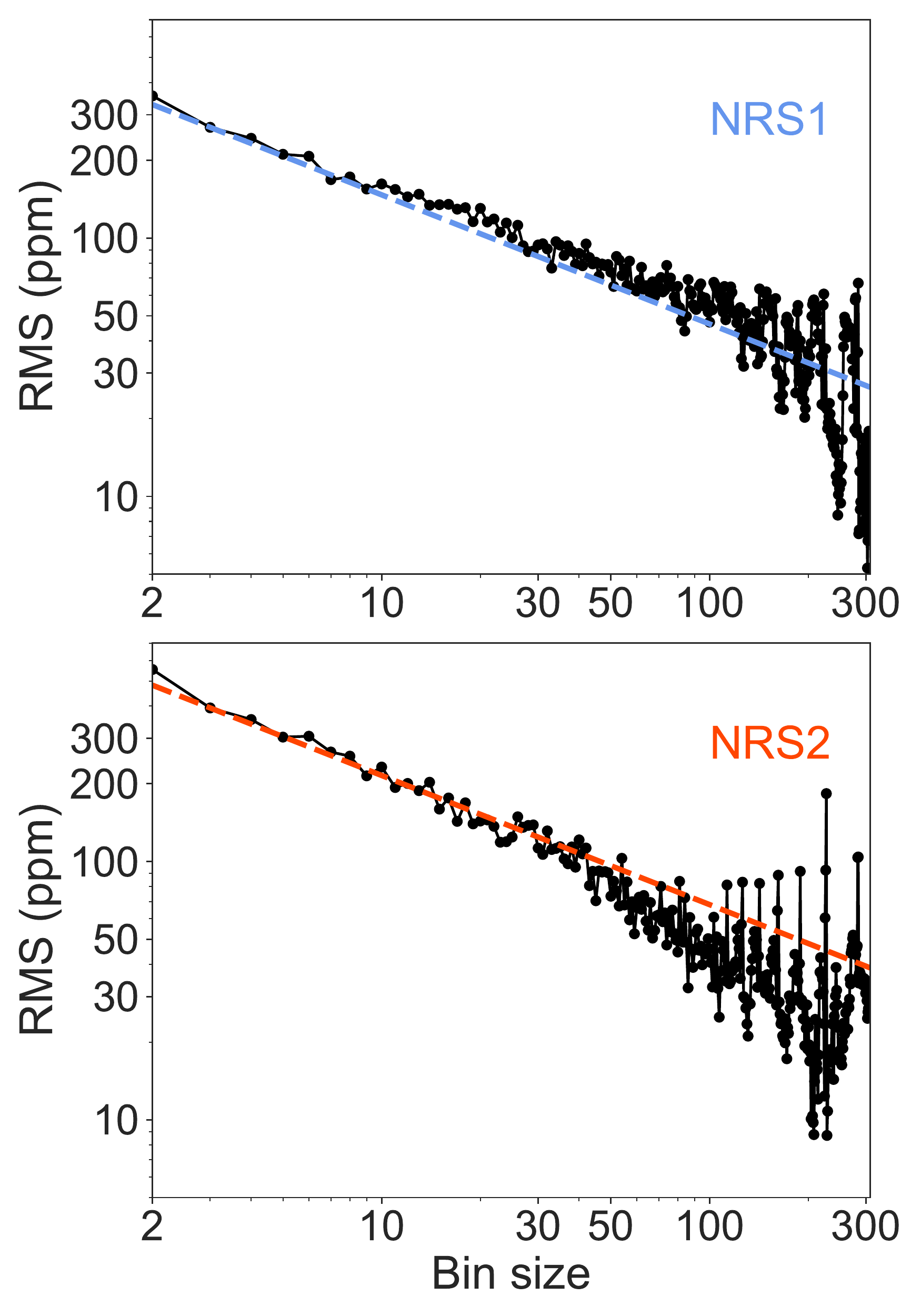}
\caption{Allan Variance plots for the band-integrated light curve fits residuals shown in Figure \ref{fig:transit} for NRS1 (top) and NRS2 (bottom). Dashed colored lines show the expected decrease on the scatters as a function of binning for perfect white-noise, i.e., 
decreasing as the square-root of the bin-size, which is a very good match to the 
actual observed data, suggesting a lack of strong correlated noise structure.}\label{fig:allan}
\end{figure}

As can be observed, the simple model defined above resulted in an excellent fit 
to our JWST/NIRSpec G395H data. This is quite impressive, considering we did not remove any 
data points for the fit, in particular data points at the beginning of the exposure. This suggests that there 
are negligible persistence effects in the NRS1 and NRS2 detectors, at least at the fluences probed by 
our observations. The observed scatter in the light curve, however, 
is larger than expected from pure read-noise and photon-noise statistics as 
calculated by the JWST pipeline by a factor of about 3; the most likely 
explanation for this is residual 1/f spatial covariance which causes extra white-noise scatter in a TSO --- noise that is not seen nor impacts the analyses 
if one performs light curve analyses at the resolution level of the instrument 
(see Section \ref{sec:wav-light-analysis} for a discussion on this), where we reach photon and read-noise limited precision. Indeed, 
the overall residual structure of 
a simple light curve fit like this shows a lack of correlated noise structure. 
We demonstrate this using Allan variance plots on the residuals, presented in Figure 
\ref{fig:allan}. The plots show that as we bin the data in time, the 
standard deviation decreases closely following a $1/\sqrt{N_{bin}}$ shape (dashed
 lines in this figure), which is consistent with the expectation for uncorrelated white-noise. 

\begin{figure}
\centering
\includegraphics[width=3.3in]{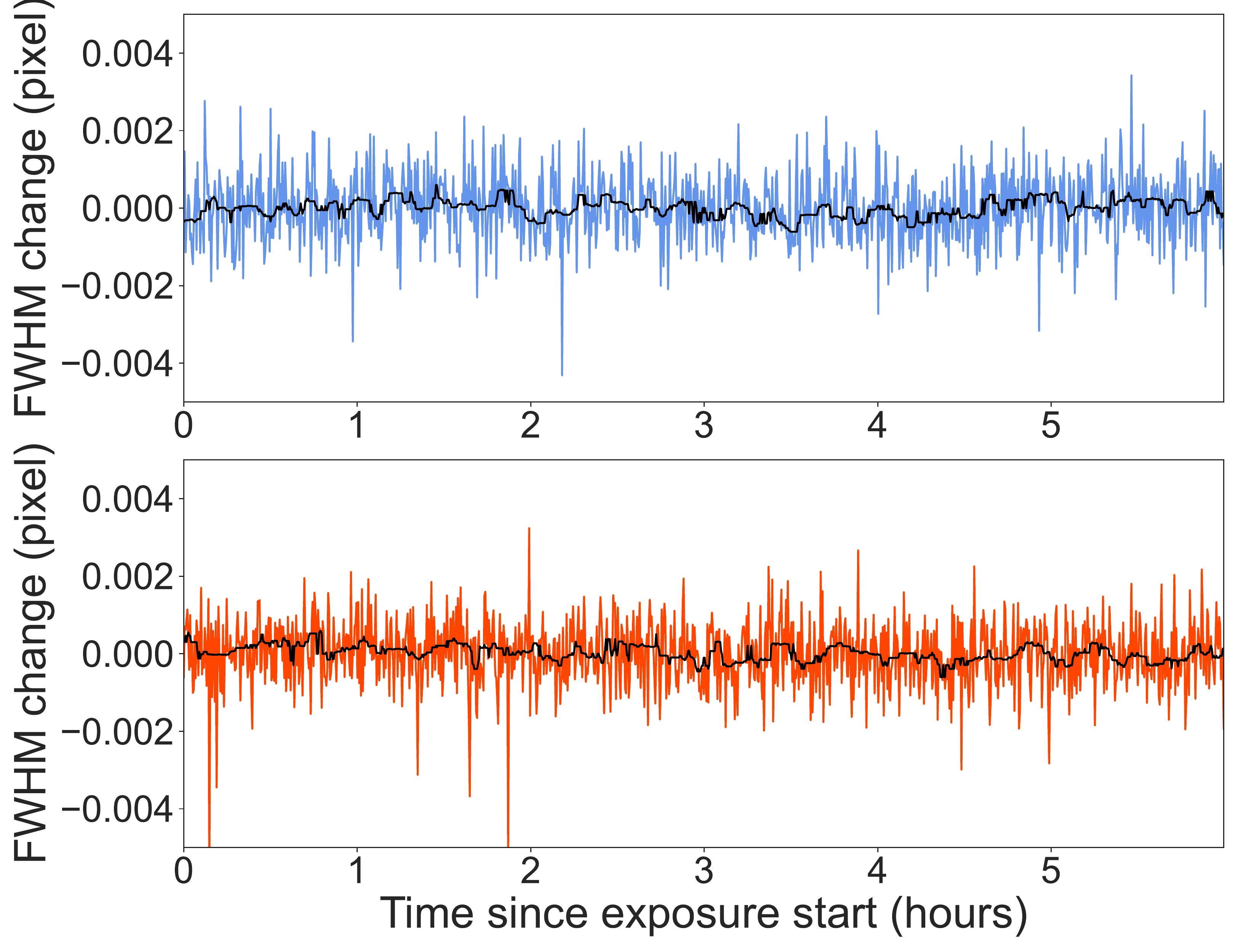}
\caption{FWHM change over the course of the exposure for NRS1 (top, blue) and 
NRS2 (bottom, red) portions of the spectral trace. Black line is a median filter with a window of 21 points 
through the measurements. Note the remarkable stability of the 
FWHM over the course of the observations --- it stays stable to about 1/1000th 
of a pixel.}\label{fig:fwhm}
\end{figure}

It is interesting to note, however, that the NRS1 transit light curve clearly shows a more pronounced exposure-long slope than the NRS2 one (about $-140$ ppm per hour for NRS1, 
compared to $40$ ppm per hour for NRS2). While we observe 
some wavelength-dependence of the slope for NRS1 (see next sub-section), the smaller slope seen in 
the NRS2 light curves is fairly constant for all wavelengths. Given that the NRS1 fluence levels vary 
relatively little at the high signal-to-noise ratio portions of the spectra as compared to NRS2 (see 
Figure \ref{fig:2dspectrum}), it is unlikely this is purely a fluence-dependent effect. Also, given the significant 
difference between the slopes on NRS1 and NRS2, it is unlikely this is, e.g., an astrophysical, observatory 
\citep[e.g., ``tilt" events;][]{rigby} and/or instrument-level effect --- all those explanations should give rise to smooth transitions from one detector to the other as well as similar effects on both detectors. The latter two set of hypotheses are also unlikely 
given the stability of both the traces (Figure \ref{fig:traces}) and the full-width at half maximum (FWHM; a detailed analysis of the FWHM shape and its time stability is presented in Appendixes \ref{appendix:trace} and \ref{appendix:fwhm}) of the spectral profile measured during the observations, which we present on Figure \ref{fig:fwhm} and which have been shown by Schlawin et al. (in prep.) and Beatty et al. (in. prep.) to be excellent tracers of ``tilt events". Given all this, it is possible the effect is due to some 
underlying detector-level effect, but this is currently under investigation. A discussion on possible causes is presented in Section \ref{sec:discussion}. 

\subsection{Wavelength-dependent light curve analysis}
\label{sec:wav-light-analysis}

\subsubsection{Light curve scatter}
\label{sec:lc-scatter}

Before jumping into the analysis of the transit event as a function of wavelength, 
we perform simple analyses to understand whether the 
light curve scatter at the native resolution of the instrument is, indeed, being correctly estimated by the JWST 
pipeline. This latter calculation includes both photon-noise and read-noise characteristics of the detector, 
including the impact of 1/f noise on a single pixel, but not the covariance this might have with nearby pixels. 

\begin{figure}
\centering
\includegraphics[width=3.4in]{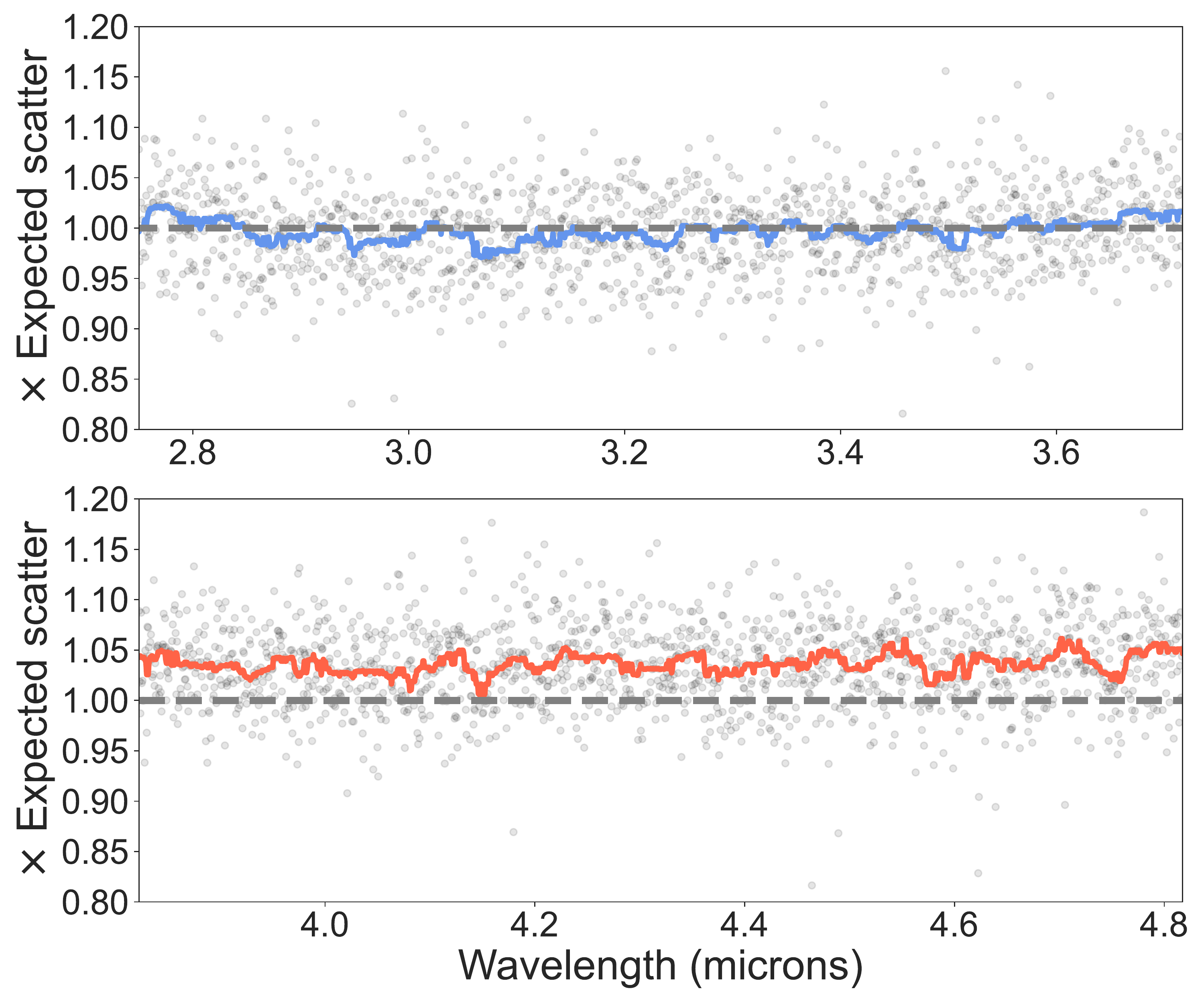}
\caption{Ratio between measured over expected (from the JWST pipeline) out-of-transit light curve scatter for our 
HAT-P-14 TSO as seen by NRS1 (top) and NRS2 (bottom). Note how the pipeline precisely predicts the observed scatter for NRS1, whereas for NRS2 there is a 
slight extra scatter at the 5\% level. Colored lines show median filters for easier visualization; grey dashed lines represent the ideal value of 1.}\label{fig:scatter}
\end{figure}

To perform this comparison, we took the out-of-transit scatter (i.e., the standard deviation of the flux time-series) on the wavelength-dependant light curves of both NRS1 
and NRS2 at the spectral sampling of the instrument (i.e., at each column 
in Figure \ref{fig:2dspectrum}) and compared that to the out-of-transit median errorbars reported by the JWST pipeline (i.e., adding in quadrature the pipeline-reported uncertainties on a given column, and taking its square-root). We then simply took 
the ratio of these two numbers, which in an ideal world should be exactly 1. These ratios are presented in 
Figure \ref{fig:scatter}.

As can be seen from Figure \ref{fig:scatter}, at the native resolution of the instrument, 
the JWST pipeline precisely predicts the noise levels observed on the out-of-transit 
light curves. There seems to be, however, a slight but significant extra scatter of about 5\% in 
the actual data when compared against the pipeline products for NRS2. The source of 
this discrepancy is currently under investigation.

Next, we perform the same analyses but performing \textit{binning} of the spectral 
channels. This is a somewhat standard procedure in transiting exoplanet spectroscopy 
for which combining the light from different wavelength bins (i.e., columns in Figure 
\ref{fig:2dspectrum}) helps to boost the signal-to-noise ratio of the light curves, 
with the band-integrated light curves in Section \ref{sec:white-light-analysis} representing the limiting 
case for this approach. The motivation for doing such an experiment on \emph{JWST} 
data is 1/f noise. While we performed some corrections on this effect in Section 
\ref{sec:1f}, these are most definitely not correcting the effect completely. 
Column-by-column median subtraction, in particular, can be roughly thought of as 
subtracting the median every 32 samples of a 1/f time-series. This will procedure thus 
will remove some but not all the covariance both within and between columns in a given integration. Hence, 
some residual covariance due to 1/f noise is expected to ``leak" into our measurements.

\begin{figure}
\centering
\includegraphics[width=3.3in]{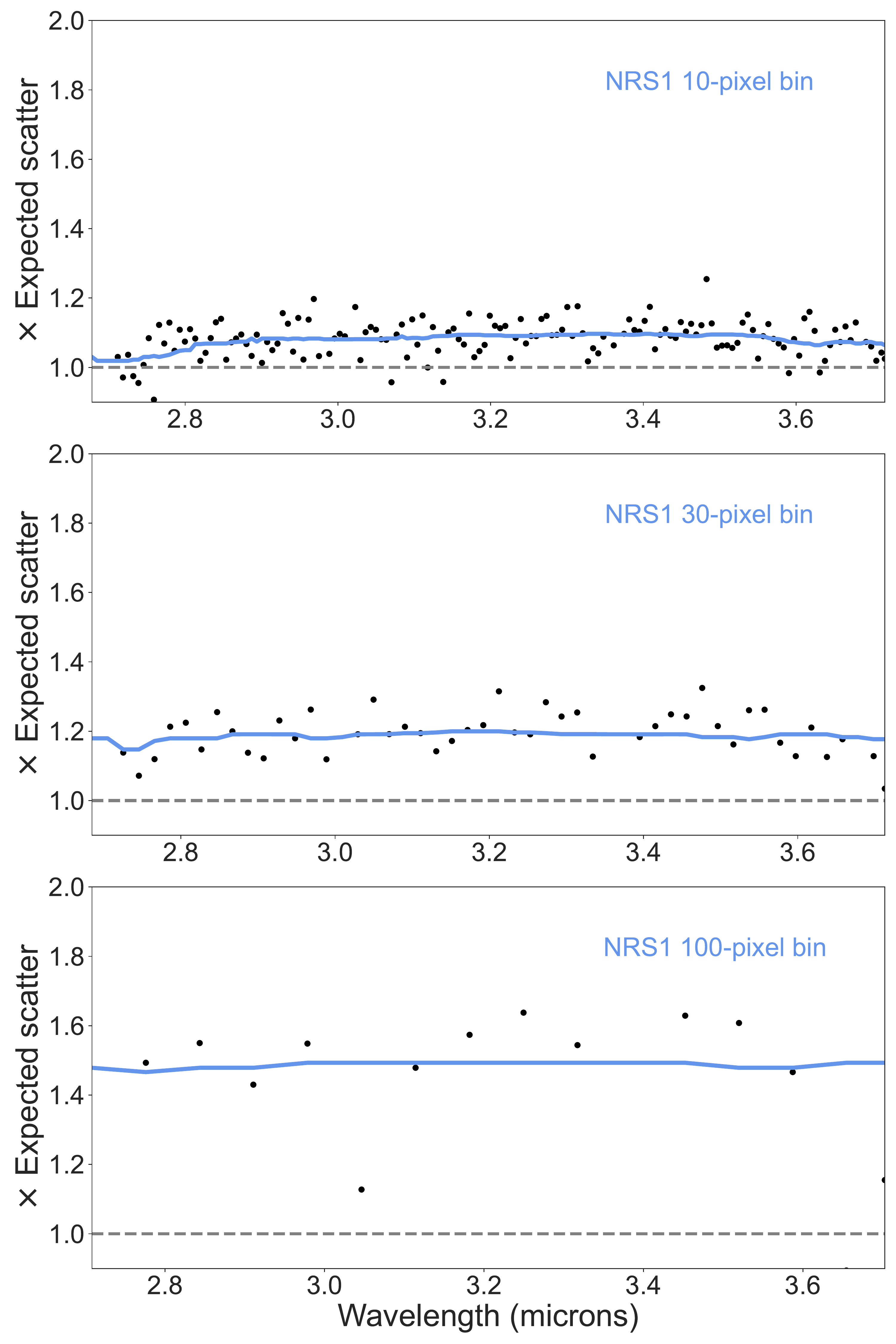}
\caption{Ratio between measured over expected (from the JWST pipeline) 
out-of-transit light curve scatter for our HAT-P-14 TSO as seen by NRS1 
using different number of spectral bins: 10 (top), 30 (middle) and 100 pixels/elements 
(bottom) panels (black points; blue lines are median filters shown for illustration). As can be seen, the larger the bin, the larger the 
deviation of the observed scatter versus the predicted one just from 
the JWST pipeline error bars. The main suspect for this behavior is 1/f noise (see text).}\label{fig:scatter-bin}
\end{figure}

Figure \ref{fig:scatter-bin} shows the same experiment as Figure \ref{fig:scatter}, but 
after performing spectral binning after spectral extraction of three different widths: 
10, 30 and 100 columns/pixels for 
NRS1 (as a comparison, the band-integrated light curve for NRS1 implies a 1543-column bin). 
The results of this experiment shows that one observes a larger scatter than the one 
one would expect by simply adding in quadrature the uncertainties reported by the JWST 
pipeline. We believe part of this is  related to the residual covariance between pixels of 
different columns described above, which makes our simple addition-of-variances to 
estimate the resulting noise of the bin a lower limit on the actual total variance. Part of it 
could also be instrumental systematics that only become evident once a better signal-to-noise 
ratio is attained via the spectral binning. A full investigation and implementation of the 
covariance due to, e.g., 1/f noise on the JWST pipeline is outside the scope of this work 
\citep[but see, e.g.,][]{schlawin}{}, as well as a detailed investigation of more subtle 
sources of systematic noise. Despite of this, in our experiments we have found that the 
best way to work with \emph{JWST} data from NIR detectors is to work at the spectral sampling 
level of the instrument (i.e., at the column-to-column level). Then, observables such as, e.g., 
the transit depth as a function of wavelength can be binned in a post-processing stage. We have 
found this methodology to give the most accurate and precise results during commissioning.

\subsubsection{Wavelength-dependent light curve modelling}
\label{sec:lc-modelling}

Following the procedures described above, we proceeded to model the wavelength-dependant 
light curves in the very same way as described for the band-integrated light curve in 
Section \ref{sec:white-light-analysis}. The only difference was that we (1) fixed the 
orbital parameters to the same ones ingested as priors on the band-integrated lightcurve analysis, (2) 
fixed the time-of-transit center to the one found in our band-integrated lightcurve analysis and (3) used priors on the square-root limb-darkening coefficients instead of letting them go free 
in the light curve fitting procedure. The prior was a truncated Gaussian defined between 0 and 
1 for both $q_1$ and $q_2$; the mean of those Gaussians was obtained by using 
the \texttt{limb-darkening} package described in \cite{espinoza}, for which we ingested the 
stellar properties of HAT-P-14 defined in \cite{bonomo} and used ATLAS \citep{atlas} intensity profiles 
to compute model limb-darkening coefficients. Then, we passed the resulting non-linear 
limb-darkening coefficients through the SPAM algorithm of \cite{spam} to obtain the final 
square-root law limb-darkening coefficients we set for the mean of our priors. The standard 
deviation of the prior for each coefficient was set to 0.1. Following 
the lessons learned in Section \ref{sec:lc-scatter}, we work at the resolution level of 
the instrument and fit each transit light curve with a combination of a transit model 
and a slope, which we found gave very good results.

\begin{figure*}
\centering
\includegraphics[width=3.5in]{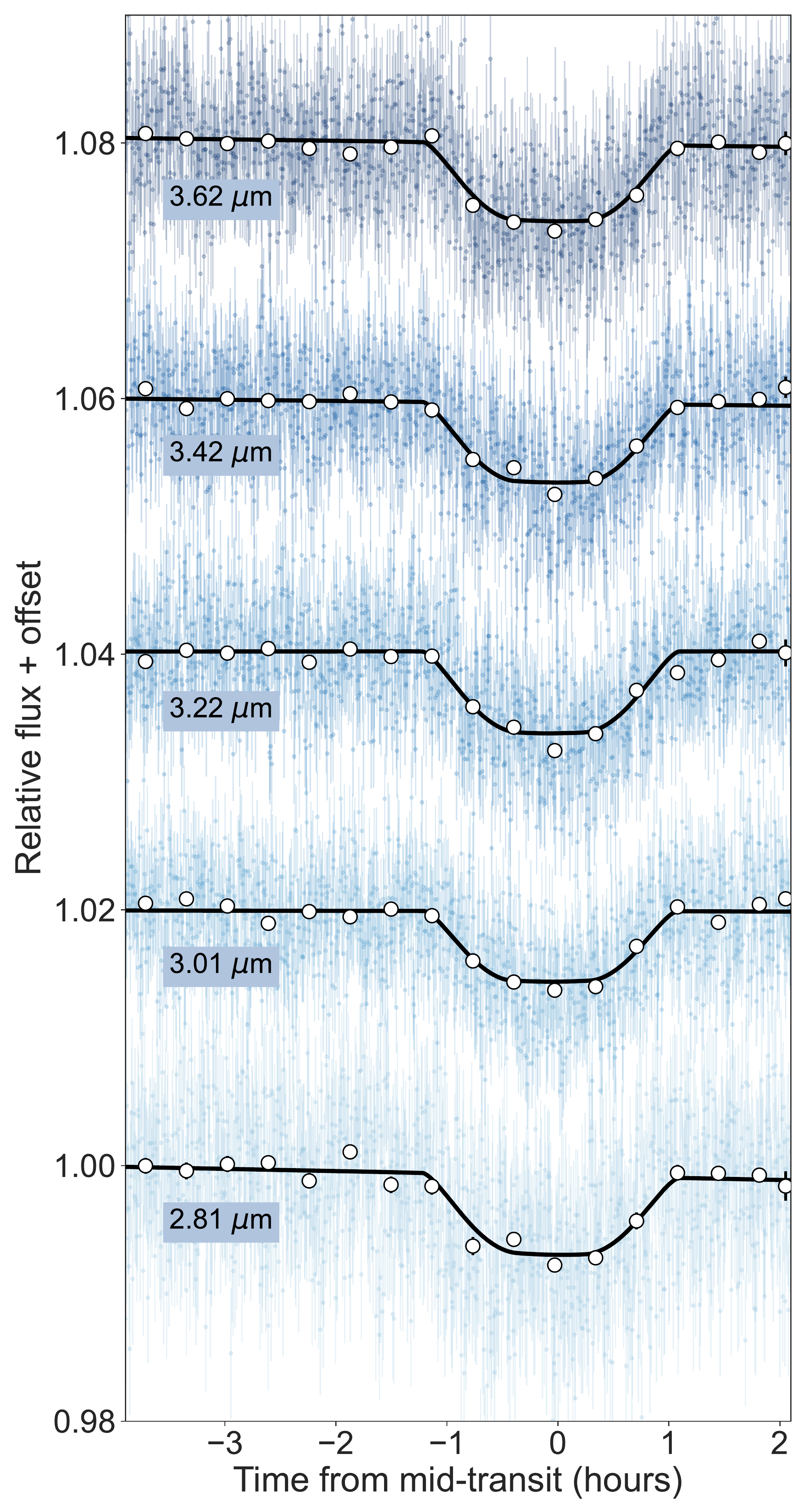}
\includegraphics[width=3.5in]{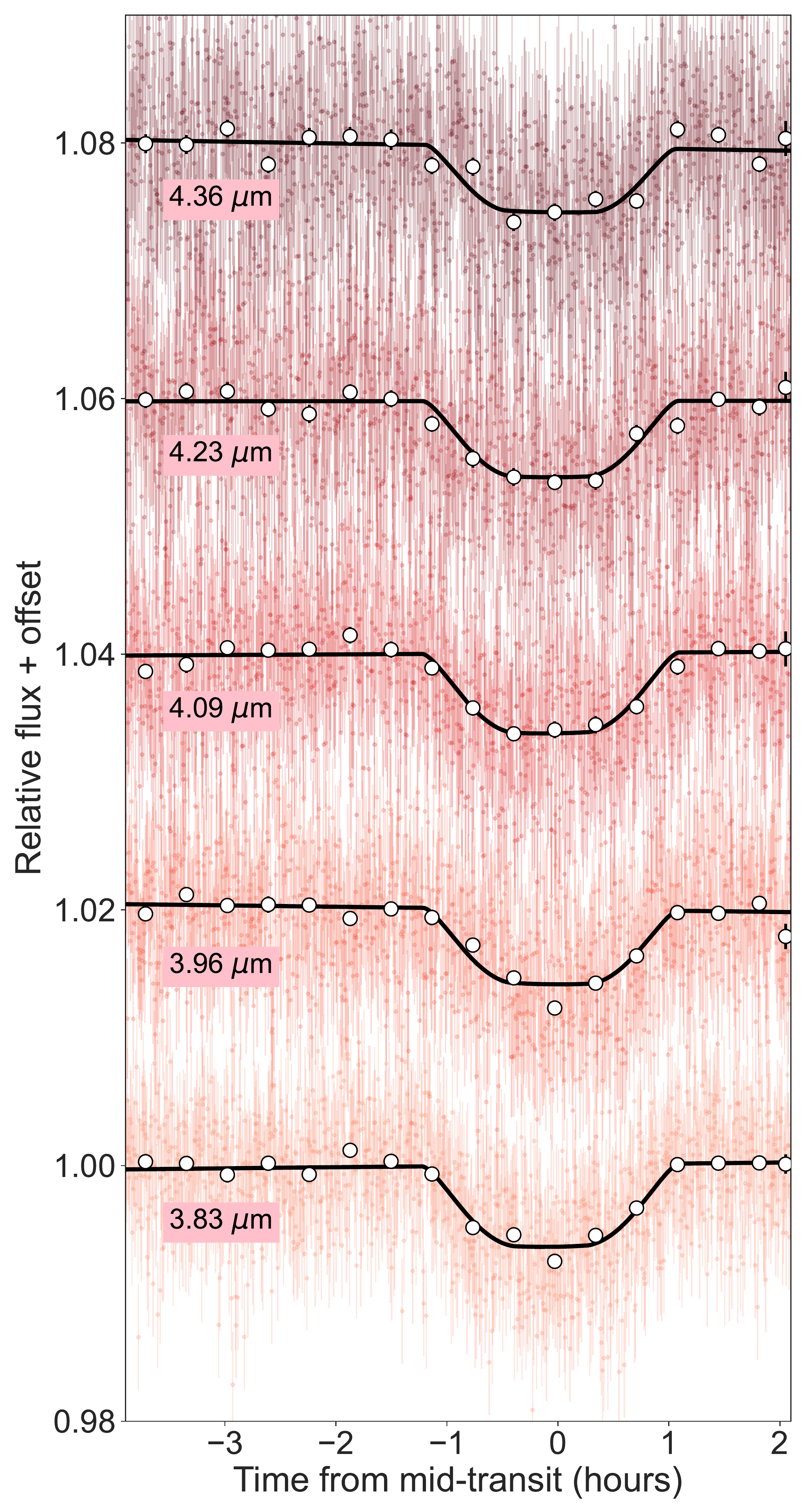}
\caption{Sample, unbinned wavelength-dependant transits of HAT-P-14~b obtained in our commissioning NIRSpec/BOTS G395H observations for NRS1 (blue, left plot) and NRS2 (red, right plot). The black solid lines show our transit plus systematic model; white points are 70-integration bins shown for illustration purposes.}\label{fig:wav-transit}
\end{figure*}

Figure \ref{fig:wav-transit} show some sample light curves fitted following our procedures. 
As can be seen, the quality of these native-resolution light curves is very good, and our 
transit model plus linear trend seems to be enough to model most of the systematic trends 
observed in the data. 

Our wavelength-dependant transits allowed us to measure two observables. First, it allowed 
us to explore the wavelength-dependence of the exposure-long linear trends already described 
in Section \ref{sec:white-light-analysis}, but it also allowed us to obtain the main 
observable used to commission the instrument: the transit depth (and its uncertainty) as a 
function of wavelength. We discuss our analyses of those two observables in the next sub-sections.

\subsubsection{Linear trend wavelength dependence}

\begin{figure}
\centering
\includegraphics[width=3.2in]{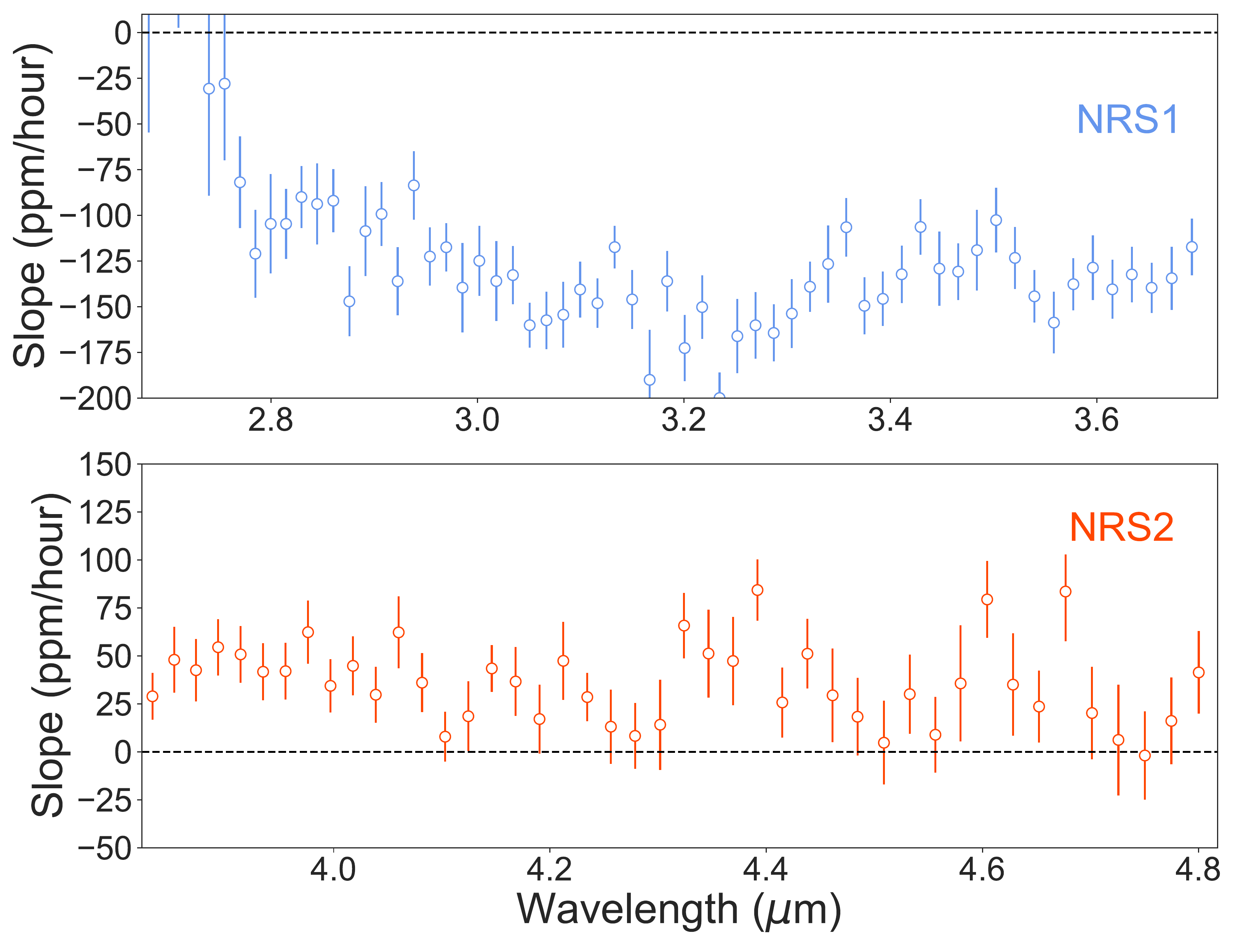}
\caption{Exposure-long slopes observed in our NIRSpec/BOTS observations for both NRS1 (top) and 
NRS2 (bottom) in ppm per hour as a function of wavelength. Dashed line indicates a no-slope scenario in both cases. Note the different values of the slopes between the two cases: negative decreasing to lower values as wavelength increases for NRS1; mostly constant positive value for NRS2. No smooth transition is seen as a function of wavelength, which suggests this is a detector-level systematic.}\label{fig:slope-wav}
\end{figure}

From our light curve modelling described in Sections \ref{sec:white-light-analysis} and 
\ref{sec:lc-modelling}, we were able to retrieve the wavelength-dependence of the linear 
trend observed both in the NRS1 and NRS2 data. While we obtained these slopes at the resolution 
element of the instrument, we bin those down to a resolution of $R=200$ for easier visualization. 
Figure \ref{fig:slope-wav} show the slopes as a function of wavelength for both NRS1 and NRS2. As 
can be observed, a slope is observed in both detectors, but the observed slope is much stronger on 
NRS1 than in NRS2 as already hinted in Section \ref{sec:white-light-analysis}.

Our results for the wavelength-dependence of the exposure-long slope are very interesting in 
particular for NRS1. First, it seems at the shortest wavelengths of the NRS1 data the slope strongly 
decreases in amplitude down to around zero. However, it quickly settles to a value of about -140 ppm/hour 
at about 3 microns, and stays more or less constant at the redder NRS1 wavelengths. The constant 
value observed for the NRS2 detector, however, is of about 30 ppm/hour --- a factor of almost 5 smaller 
in amplitude (and different sign). Given the completely different level of slopes observed between 
the detectors, the stability of other metrics (e.g., trace positions, FWHM), and the fact that the 
amplitude and sign of the slopes don't smoothly evolve as a function of wavelength from one detector 
to the other as a function of wavelength, there is a strong suggestion that the slope is somehow 
introduced at the detector level in the NRS1 detector. We further discuss this possibility in 
Section \ref{sec:discussion}.

\subsubsection{HAT-P-14~b's transmission spectrum}

\begin{figure*}
\centering
\includegraphics[width=7.5in]{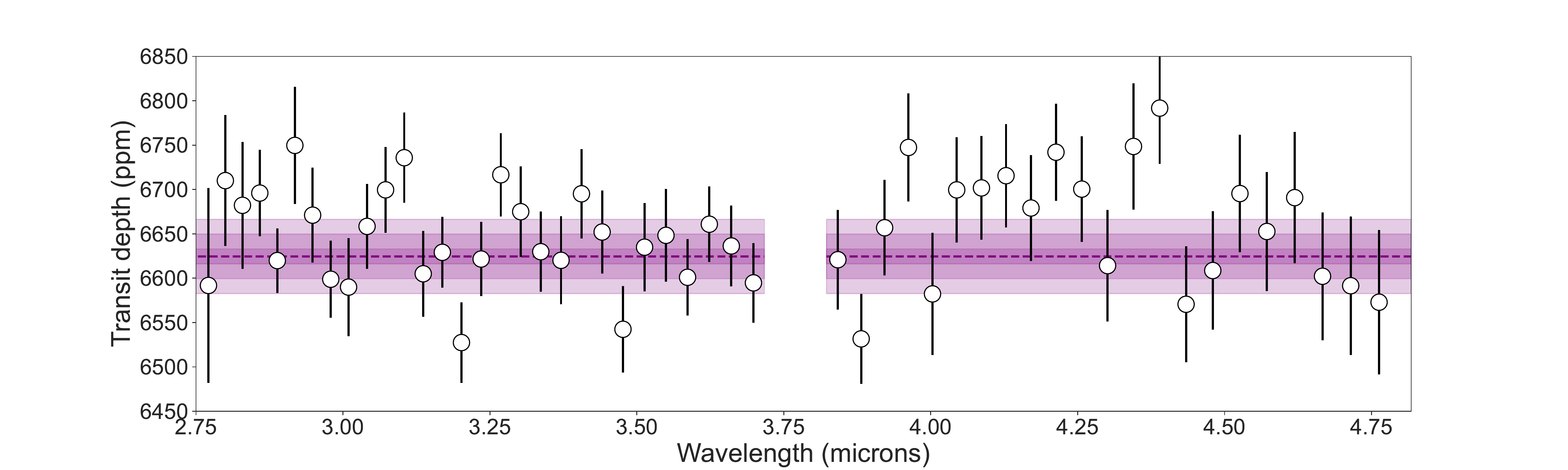}
\caption{Transmission spectrum for HAT-P-14~b as obtained by our NIRSpec/BOTS G395H observations. The high-resolution spectrum has been binned to a resolution of $R=100$ from 2.7 to 4.8 $\mu m$ (white points with error bars); the weighted mean transit depth across the bandpass is $6627 \pm 8$ ppm (purple line with bands denoting 1, 3 and 5 sigma bands around this value). The gap in the middle, at about 3.75 $\mu$m, is due to the detector gap between NRS1 and NRS2 also seen in 
Figures \ref{fig:2dspectrum} and \ref{fig:1dspectrum}. The spectrum is featureless (p-value $> 0.1$), 
as expected for a massive exoplanet like HAT-P-14~b, down to the precision at which we 
measure the transit spectrum, which is of 50-60 ppm at this resolution (see Table \ref{tab:tspec}).}\label{fig:tspec}
\end{figure*}

\begin{deluxetable}{ccc}
\tablewidth{0pc} 
\tablecaption{
    Transmission spectrum of HAT-P-14~b (at $R=100$) as measured by our NIRSpec/BOTS G395H observations. \label{tab:tspec}
}
\tablehead{
    \colhead{Wavelength} &
    \colhead{Transit Depth} & 
    \colhead{Error}\\ 
    \colhead{($\mu$m)} &  
    \colhead{(ppm)} &  
    \colhead{(ppm)}\\ 
} 
\startdata 
2.771 & 6592 & 110 \\
2.800 & 6710 & 74 \\
2.829 & 6682 & 72 \\
2.859 & 6696 & 49 \\
2.888 & 6620 & 36 \\
2.918 & 6750 & 66 \\
2.948 & 6671 & 53 \\
2.979 & 6599 & 43 \\
3.010 & 6590 & 55 \\
 $\cdot \cdot \cdot$ & $\cdot \cdot \cdot$  & $\cdot \cdot \cdot$ \\
4.526 & 6695 & 66 \\
4.572 & 6653 & 67 \\
4.619 & 6691 & 74 \\
4.667 & 6602 & 72 \\
4.715 & 6591 & 78 \\
4.763 & 6573 & 81 \\
\enddata
\tablecomments{
    A sample of the dataset is shown here. The entirety of this 
    table is available in a machine-readable form in the online journal.
}
\end{deluxetable}

Next, we retrieve HAT-P-14b's transmission spectrum from our light curve fits. 
As discussed above, the transmission spectrum we forge is obtained at the resolution 
level of the instrument, which we then bin down to a lower resolution of $R=100$. 
Figure \ref{fig:tspec} (and Table \ref{tab:tspec}) shows the transmission spectrum of HAT-P-14~b. To test 
whether this spectrum is consistent with a featureless spectrum (which is 
the expectation), we performed a chi-square test under the assumption of a 
flat spectrum and found a p-value of $0.13$ --- consistent with the featureless 
spectrum scenario. Running the test separately on NRS1 and NRS2 yielded p-values of 
$0.11$ and $0.30$, respectively --- also consistent with the featureless spectrum scenario. 
It is also important to note that all data points above 2.85 $\mu$m at $R=100$ have 
uncertainties below 100 ppm --- in particular, the median error bar 
of the spectrum on NRS1 at those wavelengths is 48 ppm, with the scatter on the 
spectrum itself being 48 ppm (i..e, both consistent with each other). Similarly, for NRS2, the median uncertainties were of 63 ppm with an actual measured standard deviation on the spectrum of 64 ppm (again, 
both consistent with each other). As we will show in Section \ref{sec:discussion}, while these precisions are about 20\% larger than what tools like PandExo \citep{PandExo} predict, they are about $\sim 10\%$ \textit{better} than pre-flight expectations from Pandeia \citep{Pandeia} if one accounts for the fact that we are fitting for a limb-darkened star and not a flat-bottomed transit as the one PandExo assumes.

One final notable result from this transit spectrum is the exquisite precision on the broadband transit depth we are able to obtain by calculating the weighted mean of the transit depth at all wavelengths using this transit spectrum: we obtain a weighted mean transit depth of $6627 \pm 8$ ppm (purple band in Figure \ref{fig:tspec}), which is in excellent agreement (within 1-sigma) of the one obtained by the \textit{JWST/NIRCam} short-wavelength photometry observations of HAT-P-14~b (Schlawin et al., in prep.).

\section{Discussion}
\label{sec:discussion}

\subsection{General TSO performance of NIRSpec/BOTS}
Via a detailed analysis of the TSO targeting a transit event of the 
exoplanet HAT-P-14~b, we have shown in Section \ref{sec:tso} the excellent 
sensitivity and stability of NIRSpec/BOTS to perform precise relative flux 
measurements, successfully enabling us to understand the two key properties 
we set as objectives for our commissioning observations: that the instrument 
is properly working for these kind of scientific objectives and that, indeed, 
the instrument can be calibrated to perform these precise flux measurements. 

\begin{figure}
\centering
\includegraphics[width=3.3in]{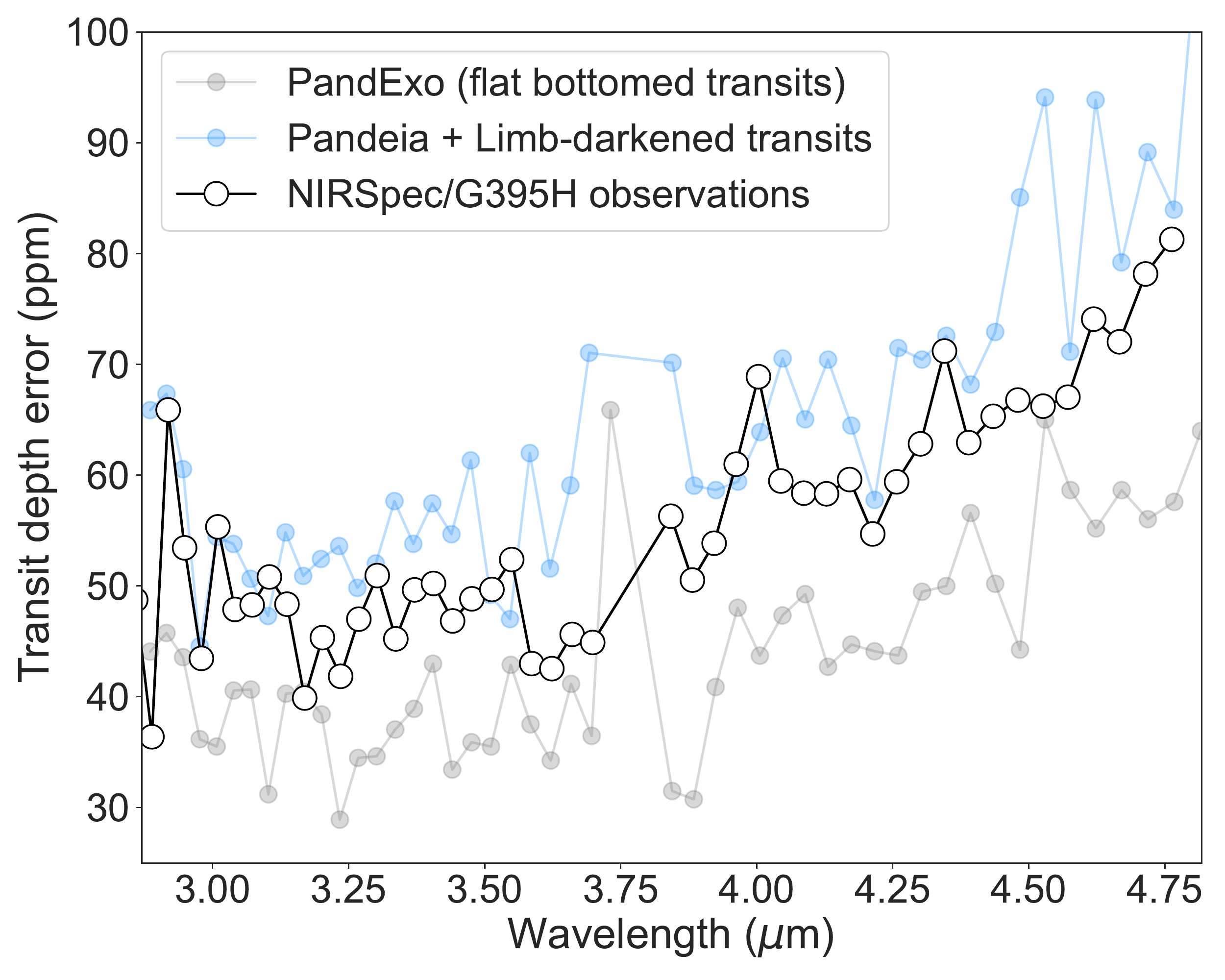}
\caption{Observed error on the transit depth at $R=100$ as a function of wavelength from our NIRSpec/G395H observations (Figure \ref{fig:tspec} \& Table \ref{tab:tspec}; white points) against pre-flight predictions from PandExo (grey points) and a simulation using Pandeia plus a limb-darkened transit (blue curve), a better 
``apples-to-apples" comparison to our observations. Our observations are about 10\% better than these latter simulations.}\label{fig:tspec-errors}
\end{figure}

The fact that the instrument is properly working for precise flux measurements 
was shown at several stages. In particular, perhaps the most straightforward is 
that, as shown in Section \ref{sec:wav-light-analysis} and Figure \ref{fig:tspec}, 
we are able to obtain a featureless transmission spectrum of the exoplanet HAT-P-14~b, 
which was the expectation given its massive nature. In addition, the precision with 
which we measure this spectrum (transit depth errors of 50-60 ppm at a resolution of 
$R=100$) is very close to what was expected prior to launch. We compare 
our results to pre-flight expectations in Figure \ref{fig:tspec-errors} via two experiments. 
The first was to compare the transit depth errors we obtain against predictions from 
PandExo \citep[][grey points in Figure \ref{fig:tspec-errors}]{PandExo}, from which 
we observe precisions that are about $\sim 20\%$ larger than those simulations. One important 
caveat of PandExo, however, is that the tool assumes flat-bottomed transits to perform its 
transit depth calculations, i.e., it omits the impact of limb-darkening in the transit 
lightcurves. While in general at the wavelengths probed by our NIRSpec/G395H observations 
this assumption might be a fair one to make, the limb-darkening effect on the transit 
lightcurve is very prominent on HAT-P-14~b due to the relatively large impact parameter of 
the transits (see Table \ref{tab:wl-params}). This can be readily observed in the ``U"-shaped 
transit lightcurve both in Figures \ref{fig:transit} and \ref{fig:wav-transit}. In order 
to perform an ``apples-to-apples" comparison to pre-flight expectations then, on our second experiment we decided to 
use the input noise information from Pandeia \citep{Pandeia} that PandExo uses to generate 
its simulations to produce noisy limb-darkened transit lightcurves instead. These simulated transit lightcurves 
had limb-darkening profiles assuming a non-linear limb-darkening law, with coefficients 
calculated as described in Section \ref{sec:wav-light-analysis} and generated 
using \texttt{batman} \citep{batman}. After simulating them at all the resolution elements sampled by NIRSpec/G395H, these were fitted in 
the exact same manner as we fitted our real transit lightcurves already 
described in Section \ref{sec:wav-light-analysis} and the resulting transit depths were binned down to $R=100$. The resulting transit depth errors from 
that experiment --- depicted as blue points in Figure \ref{fig:tspec-errors} --- show that 
our actual NIRSpec/G395H precisions are \textit{better} by about $\sim 10\%$ than pre-flight expectations. 

The fact that our results are better than pre-flight expectations in the transit 
spectrum is interesting but within expectations given the better-than-expected throughput of the \emph{JWST} NIRSpec/G395H configuration \citep{nirspec:giardino}. Furthermore, as the above experiments and the ones 
shown in Section \ref{sec:lc-scatter} show, we believe we are very close (if not 
\textit{at}) the ``photon-noise" (plus read-noise) limit, given how precisely 
the \textit{JWST} pipeline predicts the actual lightcurve scatter at the 
resolution elements sampled by the instrument. 

\subsection{Lightcurve systematics}

As it was shown in Sections \ref{sec:obs-dr} and \ref{sec:tso}, the second 
objective of our TSO observations --- to confirm that the instrument can be 
calibrated to perform precise relative flux measurements as the ones needed 
for transiting exoplanet observations --- has been 
readily met, with the level of systematic effects on the TSO itself being 
very small. However, as it was shown in Section \ref{sec:tso}, the NRS1 detector 
seems to show an exposure-long slope which had no straightforward explanation. 
While it seems this might be a detector-level effect, it's unclear what 
step and/or calibration could be producing it. A superbias, background 
and/or dark-current-related effect, for instance, would cause a source 
of dilution at the precision level of the transit spectrum, but it would be 
hard for it to produce and/or significantly change the amplitude of 
an exposure-long slope. A linearity problem could, in theory, amplify an 
existing trend in the data as well (e.g., the trend observed in the 
NRS2 detector); however, below 2.9 $\mu$m and above about 3.6 $\mu$m, NRS1 has 
a very similar fluence level to that observed in NRS2. Looking at 
Figure \ref{fig:slope-wav}, it is unlikely thus that this could be the 
effect producing the differences. In addition, an amplification and/or dilution 
source like the ones discussed would also imply the same impact on the transit 
spectrum; however, we don't see a significant difference between the transit 
spectrum obtained between NRS1 and NRS2 in Figure \ref{fig:tspec}. One possibility 
under study involves the fact that internal flat calibrations 
were performed up to two hours before the TSO exposure presented in this work. This 
could have filled some detector traps akin to those observed in, e.g., \textit{HST/WFC3} 
\citep{zhou:2017} in NRS1 which are slowly being released during the TSO. This, however, 
does not directly explain why the observed slope in NRS1 is so different from that of NRS2, despite 
the fact they are very similar detectors (with a gain difference of only about 10\%). A possibility is that while these detectors are of the same family, they do have, indeed, different inherent properties like e.g., responses to charge trapping and/or persistence. Other 
datasets obtained both with this mode and NIRSpec/Prism should be investigated in detail 
to search for the repeatability of the slopes and the corresponding amplitudes observed in our 
HAT-P-14~b observations.

\subsection{Timing accuracy of \emph{JWST} time-stamps}
\label{sec:timing}

As shown in Section \ref{sec:white-light-analysis} and presented in 
Table \ref{tab:wl-params}, our observations allowed us to obtain a precise combined time-of-transit between NRS1 and NRS2 of $2459729.706791^{+0.000094}_{-0.000093}$  BJD TDB --- an uncertaintiy on the timing of the event of only 8 seconds. While from the lessons learned in Section \ref{sec:wav-light-analysis} an even better precision is achievable on this timing if the lightcurve analyses are performed at the resolution level of the instrument, this timing precision is enough to test the accuracy of the observatory time-stamps at the tens of seconds accuracy, at least as compared to another mission: \emph{TESS}. Additional data to the one used in Section \ref{sec:white-light-analysis} was recently released by the \emph{TESS} mission for HAT-P-14~b for sectors 52 and 53 which include \textit{a contemporaneous} transit to that observed by our \emph{JWST/NIRSpec} observations presented in this work. Using the same methods as presented in \cite{patel}, we performed a joint fit of all the 2-minute cadence photometry from \emph{TESS}, including Sectors 25, 26, 52 and 53. We retrieve transit parameters from those \emph{TESS} observations consistent with those presented in Section \ref{sec:white-light-analysis} and Table \ref{tab:wl-params}, and obtain a transit ephemerides of $P = 4.6276618 \pm 0.0000018$ days and $t_0 = 2459766.72798 \pm 0.00020$ (BJD TDB). In particular, for the transit event observed by \emph{JWST/NIRSpec}, the \emph{TESS} observations observe a time-of-transit of $2459803.74927 \pm 0.00022$ BJD TDB, which is in excellent agreement with the timing of our \emph{JWST/NIRSpec} observations --- the difference between the two being $9 \pm 19$ seconds --- consistent with zero at 1-$\sigma$. 

\section{Summary}
\label{sec:summary}

We have presented spectrophotometric TSO commissioning observations of HAT-P-14~b using NIRSpec/BOTS 
onboard \emph{JWST}, which we used to enable the mode for scientific usage. 
We measured a transmission spectrum of the exoplanet down to a precision of 
50-60 ppm at $R=100$, showcasing in turn the excellent stability and sensitivity 
of this instrument/mode to perform precise transiting exoplanet spectrophotometry. 

We find the above quoted precisions are, in turn, very close to pre-flight expectations. When compared against PandExo, these precisions seem to be about 20\% larger than what 
the tool predicts, but we find most if not all of this difference is due to the assumption of a flat-bottomed transit lightcurve by the tool. If we instead simulate limb-darkened transit lightcurves using the pre-flight noise expectations from Pandeia (which is the engine both PandExo and the JWST Exposure Time Calculator\footnote{\url{https://jwst.etc.stsci.edu/}} itself uses), we find the observed transit depth errors are about $10\%$ better than those simulations. 

While the trace position and FWHM remained fairly constant during our 
observations ---down to 1/500th and 1/1000th of a pixel, respectively --- we 
did find a low-amplitude exposure-long slope on the two detectors used to capture 
the spectra of HAT-P-14. In particular, the slope observed in NRS1 is much stronger and wavelength-dependent 
than the one observed in NRS2, and all evidence points to it being a detector-level effect. Investigations to understand this trend are currently ongoing.

Through a series of experiments, we also found that 1/f detector noise, as expected from 
pre-flight experiments \citep{schlawin, zafar}, seems to be an important component to pay attention 
to when studying \emph{JWST} TSOs with NIRSpec/BOTS. In particular, it seems binning spectral channels 
causes a degradation on the signal-to-noise ratio that might be in part caused by the residual 
covariance this detector effect ingests into the lightcurves when binning. We found that the best way to 
work with \emph{JWST} data from NIR detectors such as the ones in NIRSpec/BOTS is to work at the 
spectral sampling level of the instrument (i.e., at the column-to-column level). Then, observables such 
as, e.g., the transit depth as a function of wavelength can be binned in a post-processing stage. 
This methodology gave the most accurate and precise results during our commissioning analyses.

We conclude that the two main objectives of these NIRSpec/BOTS TSOs have been met: our analysis shows that the instrument is properly working to measure the precise relative flux measurements implied by the technique of transit spectroscopy, and that indeed, the instrument can be calibrated to perform these precise measurements. JWST NIRSpec is thus ready to perform cutting edge exoplanetary science for Cycle 1 and beyond.

\begin{acknowledgments}

All figures were prepared using \texttt{Python 3.8.13} with the aid of packages \texttt{matplotlib 3.5.2}, \texttt{NumPy 1.22.4}, \texttt{SciPy 1.8.1} and 
\texttt{juliet 2.2.0}. This work is based on observations made with the NASA/ESA/CSA James Webb Space Telescope. The data were obtained from the Mikulski Archive for Space Telescopes at the Space Telescope Science Institute, which is operated by the Association of Universities for Research in Astronomy, Inc., under NASA contract NAS 5-03127 for JWST. These observations are associated with program \#01118. This paper includes data collected by the TESS mission. Funding for the TESS mission is provided by the NASA's Science Mission Directorate.

\end{acknowledgments}

\appendix
\section{Periodogram analysis of the trace, FWHM and residual flux time-series}
\label{appendix:trace}

\begin{figure*}
\centering
\includegraphics[width=7in]{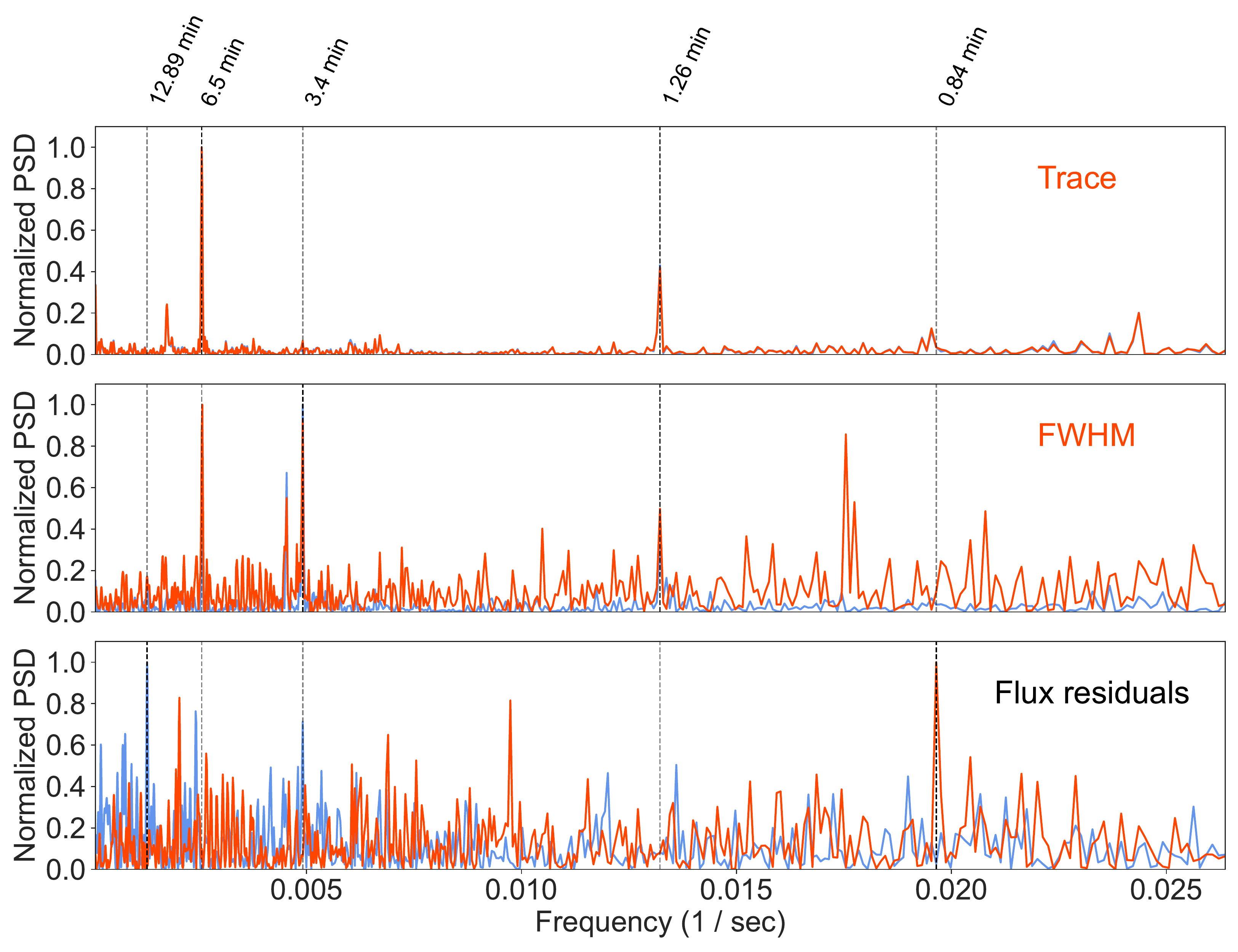}
\caption{Power Spectral Density (PSD) of the trace (top), FWHM (middle), and band-integrated light curve flux residuals (bottom) for both the NRS1 (blue) and NRS2 (red) time-series, normalized to the highest peak. The most obvious peaks accross the different PSD's are labeled with black dashed lines; the period of each of those peaks is indicated on the top panel. Note how most of the peaks in the band-integrated light curve time-series residual PSD don't really appear on the trace and the FWHM PSDs, with the exception, perhaps, of a peak at 3.4 minutes, where the FWHM also peaks. This is not a very significant peak in the band-integrated light curve PSD nonetheless (FAP $>$ 0.1).}\label{fig:psd}
\end{figure*}

As shown in Figure \ref{fig:traces}, the trace position movement is consistent between the detectors, and shows the same characteristics in time: 
an apparent break in the position about 1.5 hours since the start of the exposure, and some 
high-frequency variations. These high-frequency variations are also evident from the FWHM time-series. To characterize the features in these time-series we perform a periodogram analysis on the trace position time-series, as well as the FWHM and the residuals of the band-integrated light curve fits presented in Section \ref{sec:obs-dr}. These are presented in Figure \ref{fig:psd}.

As can be seen from Figure \ref{fig:psd}, the trace and FWHM power spectral densities (PSDs) have some peaks in common. In particular, the strongest signal at 6.5-minutes appears on both periodograms. While 
this same peak does not directly appear in the band-integrated light 
curve residual PSD (bottom panel of Figure \ref{fig:psd}) a peak in 
the NRS1 residual lightcurve \textit{does} show up at about twice 
this period (12.89 min.). The 1.26-minute peak clearly observed on the trace PSD also seems to appear on the FWHM PSD, and 
not at all in the band-integrated light curve flux residual PSD. The FWHM does appear to have an extra peak in the PSD at about 3.4 minutes which seems to 
show up in the flux residuals too, although it's not very significant 
in this later PSD (we measure a False Alarm Probability, FAP, for this peak of $>$ 0.1). The time-scales of the observed peaks in the trace and FWHM periodograms all seem consistent with the time-scales of the thermal cycling of heaters in the Integrated Science Instrument Module (ISIM) onboard JWST \citep{rigby}. Whether this is an actual \textit{causal} relationship is under investigation.

\section{FWHM variation accross the G395H trace}
\label{appendix:fwhm}

\begin{figure*}
\centering
\includegraphics[width=7.1in]{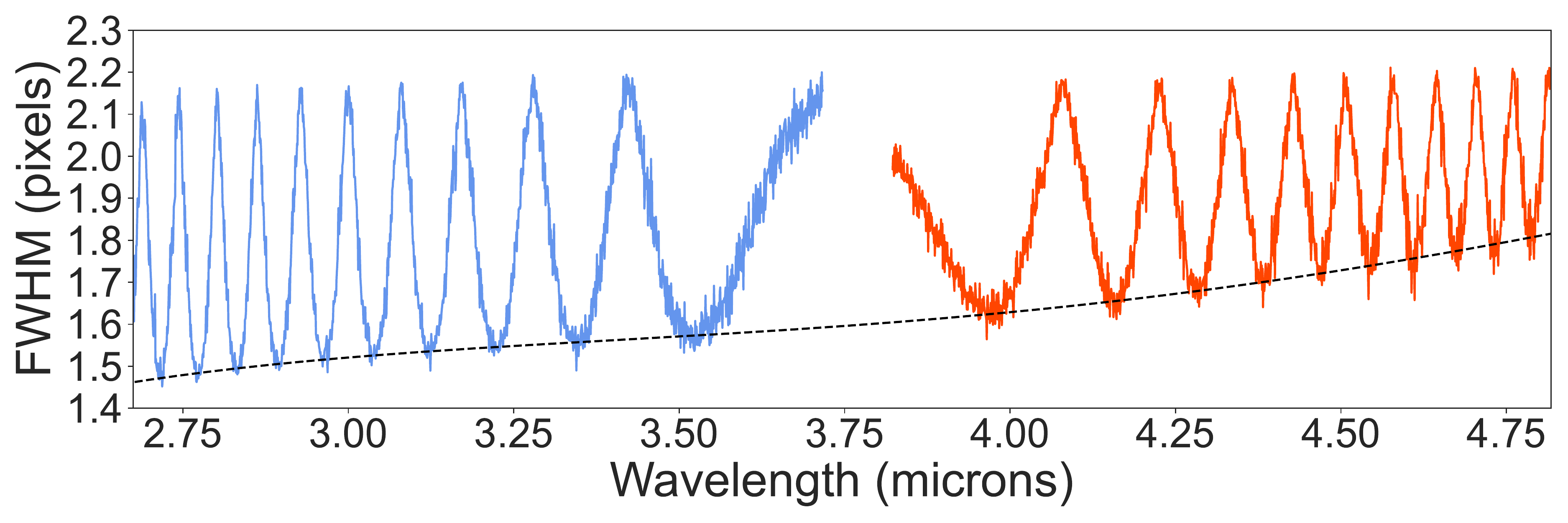}
\caption{FWHM as a function of wavelength as measured from our G395H observations from NRS1 (left, blue) and NRS2 (right, red). Note how the FWHM quickly oscillates as a function of wavelength with an amplitude of about 0.6 pixels; we believe the oscillations are almost purely a detector sampling effect and not real variations caused by the optical elements of the G395H trace. The dashed line shows the lower envelope of this FWHM shape, which should be a good estimate of the real FWHM variation as a function of wavelength of the instrument (see text for details).}\label{fig:fwhm-shape}
\end{figure*}

A notable feature to describe that was observed in our 
G395H observations is the evident variation of the FWHM as a 
function of wavelength/position in the detector, which we believe 
is almost purely a sampling effect. To obtain the FWHM at each 
position along the trace (and hence, as a function of wavelength), 
we fit a spline to the cross-dispersion profile, subtracting the half-maximum of this spline to itself, and then searching for the roots of it to find the FWHM. The median FWHM as a function of wavelength (obtained by obtaining the median FWHM across integrations) as measured by our G395H observations of HAT-P-14~b are presented in Figure \ref{fig:fwhm-shape}.

As can be observed, the FWHM seems to vary quite rapidly throughout the detector with an amplitude of about 0.6 pixels. On average, the FWHM seems to slowly increase as a function of wavelength, with the amplitude of the modulation \textit{decreasing} as a function of wavelength. Moreover, the lenghtscale at which these modulations occur seem to decrease towards the edge of the wavelength range, being larger in the middle. 

The behavior of the FWHM change across the trace is most likely a sampling effect related to the significantly tilted shape of the trace and the narrow shape of the PSF in the cross-dispersion direction, which makes the profile to be undersampled. There is ample evidence that this is the case in Figure \ref{fig:fwhm-shape} itself: the FWHM variation is much milder as a function of wavelength in the center of the wavelength range, where the trace is much less tilted than at the edges, where the variation is the strongest. We also visually inspected the cross-dispersion profiles at the peaks and valleys observed in Figure \ref{fig:fwhm-shape}. There, we observed that indeed, on the peaks, where the FWHM appears to be larger, the trace position is right at the mid-point between two pixels, whereas when the FWHM appears to be smaller the trace is almost exactly positioned in the middle of a pixel.

The above suggests thus that the best measurement of the FWHM as a function of wavelength when measured with the methods outlined above would be to measure the lower envelope of our retrieved FWHM as a function of wavelength. This envelope is presented with dashed lines in Figure \ref{fig:fwhm-shape}, and is a fit to the local minima of the FWHM as a function of wavelength with a fourth degree polynomial, which we observed gave an adequate fit to them (RMS of 0.0075 pixels):

\begin{equation*}
    \textnormal{FWHM} = c_0 + c_1 \lambda + c_2 \lambda^2 + c_3\lambda^3 + c_4\lambda^4,
\end{equation*}

with the FWHM in pixels, $\lambda$ the wavelength in microns and the coefficients being $(c_0, c_1, c_2, c_3, c_4) = (-4.6210,  6.1239, -2.2736,  0.3697, -0.0216)$. This implies the FWHM evolves from about 1.52 pixels at 3 $\mu$m to about 1.85 pixels at 5 $\mu$m for G395H.

\bibliography{manuscript}{}
\bibliographystyle{aasjournal}

\end{document}